\documentclass[conference]{IEEEtran}
\ifCLASSINFOpdf
\else
\fi
\usepackage{mathtools}
\usepackage{color}

\newcommand{\eq}[1]{
    \begin{equation}
    \begin{aligned}
    #1
    \end{aligned}
    \end{equation}
}

\newcommand{\image}[4]{
    \begin{figure}[ht]
    \centering
    \includegraphics[width=#2\textwidth]{#1}
    \caption{#3}
    \label{#4}
    \end{figure}
}

\makeatletter
\def\BState{\State\hskip-\ALG@thistlm}
\makeatother

\usepackage{algorithm}
\usepackage[noend]{algpseudocode}
\usepackage{subcaption}
\usepackage{url}
\usepackage{blindtext}
\usepackage{scrextend}
\usepackage{adjustbox}
\usepackage{graphicx}
\usepackage{cite}
\usepackage{hyperref}

\newcommand{\eat}[1]{}

\newenvironment{myitemize}
{
    \begin{list}{\labelitemi}{\leftmargin=1em}
        \setlength{\topsep}{0pt}
        \setlength{\parskip}{0pt}
        \setlength{\partopsep}{0pt}
        \setlength{\parsep}{0pt}         
        \setlength{\itemsep}{0pt} 
}
{
    \end{list} 
}

\begin{document}
%
\title{Elastic Resource Management with Adaptive State Space Partitioning of Markov Decision Processes}


%

\author{\IEEEauthorblockN{Konstantinos Lolos, Ioannis Konstantinou, Verena Kantere and Nectarios Koziris}
\IEEEauthorblockA{School of Electrical and Computer Engineering, 
National Technical University of Athens\\
Email: \{klolos, ikons, nkoziris\}@cslab.ece.ntua.gr, verena@dblab.ece.ntua.gr
}}


%


\maketitle

\begin{abstract} \label{sec:abstract}
%
Modern large-scale computing deployments consist of complex applications running over machine clusters. An important issue in these is the offering of elasticity, i.e., the dynamic allocation of resources to applications to meet fluctuating workload demands. Threshold based approaches are typically employed, yet they are difficult to configure and optimize. Approaches based on reinforcement learning have been proposed, but they require a large number of states in order to model complex application behavior. Methods that adaptively partition the state space have been proposed, but their partitioning criteria and strategies are sub-optimal. In this work we present \emph{MDP\_DT}, a novel full-model based reinforcement learning algorithm for elastic resource management that employs adaptive state space partitioning. We propose two novel statistical criteria and three strategies and we experimentally prove that they correctly decide both where and when to partition, outperforming existing approaches. We experimentally evaluate \emph{MDP\_DT} in a real large scale cluster over variable not-encountered workloads and we show that it takes more informed decisions compared to static and model-free approaches, while requiring a minimal amount of training data.
%

\end{abstract}



%
\IEEEpeerreviewmaketitle

\section{Introduction}\label{sec:introduction}

Modern large-scale computing environments, like large private clusters, cloud providers and data centers may have deployed tenths of platforms, like NoSQL and traditional SQL database servers, web servers, etc on thousands of machines, and run on them hundreds of services and applications \cite{verma2015large}. Moreover, the infrastructure architecture and system configuration may vary significantly among such environments, but also within the same environment. For example, it is common for such environments to have local or distributed storage, and run applications on VMs or bare metal. 

A vital issue in such environments is the allocation of resources to platforms and applications so that they are neither over-provisioned, nor under-provisioned, aiming to avoid both resource saturation and idling, and having as utmost goal fast execution of user workload while keeping the cost of operating the infrastructure as low as possible.

Managing the above trade-off and achieve a truly elastic behavior is quite challenging for multiple reasons. First, the number of system and application parameters that affect behavior (i.e., performance) is exceedingly large; therefore, the number of possible states of the system, which correspond to combinations of different values for all such parameters is exponentially large. 
Facebook for instance deal with this complexity with a proprietary highly sophisticated distributed system designed only for configuration management \cite{tang2015holistic}, whereas Google's Borg \cite{verma2015large} manages hundreds of thousands of jobs deployed in tens of thousands of machines.


 Second, the value range or the interesting values of such parameters may not be known; moreover, many of these parameters are continuous instead of discrete (e.g. cluster and load characteristics, live performance metrics, etc.), making it necessary to devise ahead techniques for their discretization in a way that the number of discrete values is kept small, but ranges of continuous values that lead to different system behavior correspond to different discrete values. Third, most often we do not know if and how a parameter, or a parameter value set affects the system behavior. Therefore, we do not know if changing the value of this parameter will make an impact, and furthermore, a desirable impact to the system behavior. Fourth, the time interval between two consecutive resource management decisions is usually at least in the order of minutes, reducing the collection rate of training data. Nevertheless, the resource management technique should be able to work with little such data. All four challenges are hard to address even for static workloads and applications, and become insurmountable for dynamic ones. 


Since the issue of resource management in elastic environments is so vital and challenging, there are numerous efforts to address it in both the industry and research.
Public cloud providers such as Amazon, Google, Microsoft and IBM offer autoscaling services \cite{aws_autoscale, azure_autoscale, softlayer_autoscale, rackspace_autoscale, google_autoscaling}. These employ threshold-based rules or scheduled actions based on a timetable to regulate infrastructural resources (e.g., if mean CPU usage is above 40\% then add a new VM). However, such solutions do not address any of the four challenges discussed above. 
Some research approaches to this issue also explore threshold-based solutions \cite{shen2011cloudscale, lim2010automated, das2013elastras, trushkowsky2011scads, nguyen2013agile}. More sophisticated approaches employ \emph{Reinforcement Learning} (RL) algorithms such as \emph{Markov Decision Processes} (MDP) and \emph{Q-Learning}, algorithms which are natural solutions for decision making problems and offer optimality guarantees under conditions. These approaches suffer from important limitations that derive from the assumption of a priori knowledge of parameters and their role to system behavior, as well as from the curse of dimensionality as a result of their effort to create a full static model of the computing environment \cite{rao2009vconf, bu2009reinforcement, bu2013coordinated, tiramola1, tiramola2, barrett2013applying, ortiz2016perfenforce}.


In this work we address all four challenges of elastic resource management in a large-scale computing environment by employing RL in a novel manner that starts from one global state that represents the environment, and gradually partitions this into finer-grained states adaptively to the workload and the behavior of the system; this results in a state space that has coarse states for combinations of parameter values (or value ranges) for which the system has unchanged behavior and finer states for combinations for which the system has different behavior. Therefore, the proposed technique is able to zoom into regions of the state space in which the system changes behavior, and use this information to take decisions for elastic resource management. Specifically, the proposed technique works as follows:
\textbf{Adoption of a full model.}  Since decisions are taken in intervals in the order of minutes, it is realistic to maintain a full MDP model of the system. Information about the behavior of the system is acquired at a slow
rate, limiting the size of the model and making possible expensive calculations for each decision.
\textbf{Adaptive state space partitioning.} We create a novel decision tree-based algorithm, called \emph{MDP\_DT}\footnote{Available at \url{https://github.com/klolos/reinforcement_learning}} that dynamically partitions the state space when needed, as instructed by the behavior of the system.
This allows the algorithm to work on a multi-dimensional continuous state space, but also to adjust the size of the state space based on the amount of information on the system behavior. The algorithm starts with one or a few states, and the MDP model is trained with a small amount of data. As more data on the behavior of the system are acquired, the number of states dynamically increases, and with it increases the accuracy of the model.
\textbf{Splitting criteria and strategies.} The algorithm can take as input criteria for splitting the states; such criteria aim to partition the existing behavior information, with respect to the measured parameter values, into subsets that represent the same behavior. We propose two novel criteria, the \emph{Parameter test} and the \emph{Q-value}. Also, the algorithm
can adopt various splitting strategies that perform splitting of one or multiple states, employing small or big amounts of information on behavior. 
\textbf{Reuse of information on system behavior.} It is essential that we do not waste collected information.
Therefore, if an old state is replaced with two new ones, the information used to train the old state is re-used to train the two new ones. This way, even though new states are introduced, these are already trained and their values \eat{and Q-values} already represent all the experiences acquired since the start of the model's life. 


Overall the contributions of this work are:
\begin{myitemize}
\item The \emph{MDP\_DT} algorithm that performs adaptive state space partitioning for elastic resource management.
\item Two splitting criteria, the \emph{Parameter test} and the \emph{Q-value test} that decide how the system behavior is differentiated depending on the measured parameter values, and split states accordingly.
\item Three basic splitting strategies, \emph{Chain Split}, \emph{Reset Split} and \emph{Two-phase Split},  which can be used in combination depending on the computing environment to be modelled.
\item A thorough experimental study on simulation that provides insight on the behavior of the \emph{MDP\_DT} algorithm, the splitting criteria, employed statistical tests and splitting strategies, and allow the calibration of the algorithm for optimal performance.
\item An extensive experimental study on a real large-scale elastic cluster based on the calibration resulted from the simulation study, which proves the effectiveness of \emph{MDP\_DT} in taking optimal resource management decisions fast, while taking into account tenths of parameters, and the superiority of the algorithm in comparison with classical full model-based and model-free-based algorithms.
\end{myitemize}

The rest of this paper is organized as follows: 
Section \ref{sec:markov} reviews MDP, which is the theoretical basis of our approach,
and describe the advantages and disadvantages of well known algorithms within that context.
Section \ref{sec:implementation} describes the proposed \emph{MDP\_DT} algorithm as well as splitting criteria and strategies.
Section \ref{sec:simulations} presents a thorough experimental study on simulation. 
Experimental results from testing our proposal on a real cloud computing environment are detailed in 
Section \ref{sec:experimental}, related work is presented in Section \ref{sec:related} and Section \ref{sec:discussion} presents a detailed discussion on the outcomes of this work.
Finally, Section \ref{sec:conclusions} concludes the paper. 

\section{Markov Decision Processes} \label{sec:markov}
In this section we give an overview of MDPs.
An MDP consists of the following:
\begin{myitemize}
\item A set of states $\mathcal{S}$
\item A set of actions $\mathcal{A}$
\item A reward function $\mathcal{R} : \mathcal{S} \times \mathcal{A} \times \mathcal{S} \rightarrow \Re$
\item A transition function $\mathcal{T} : \mathcal{S} \times \mathcal{A} \times \mathcal{S}
      \rightarrow [0,1]$
\end{myitemize}

\image{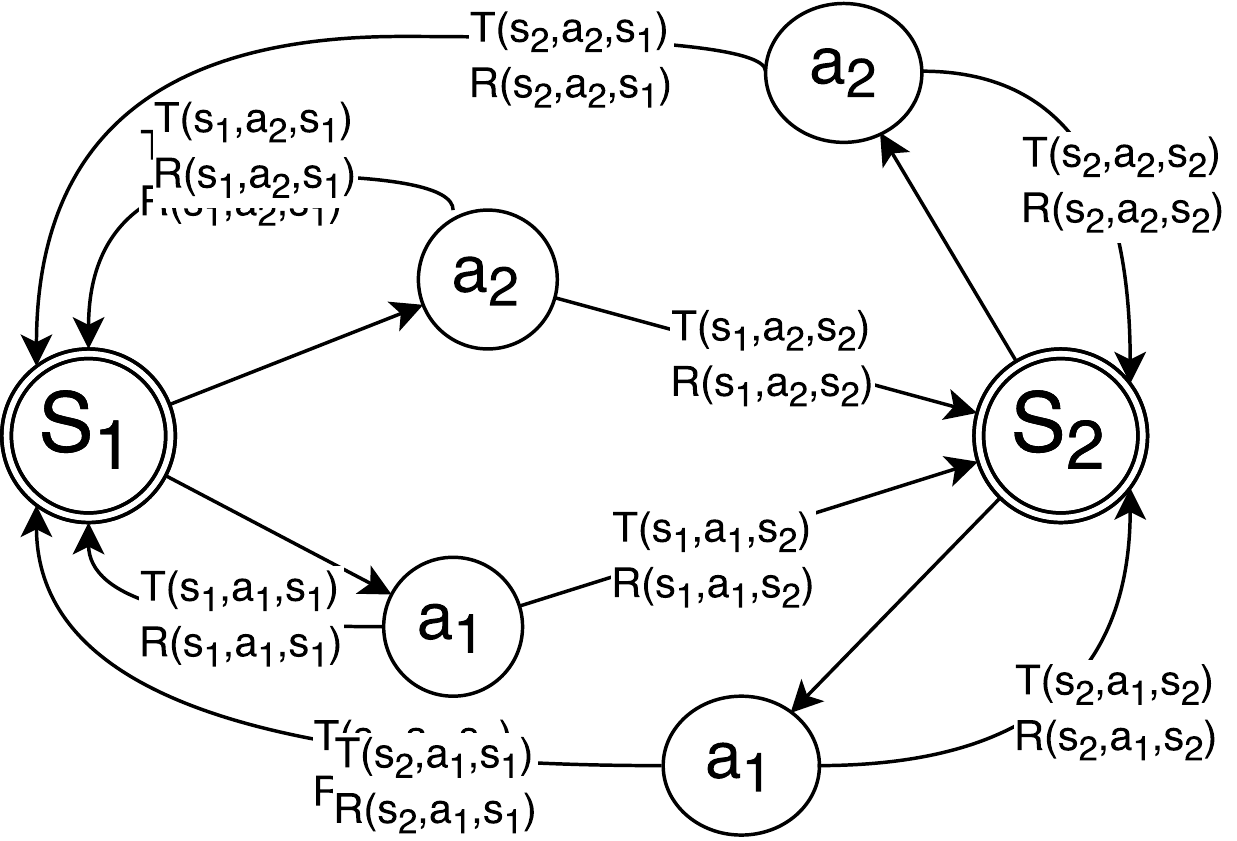}{0.3}{Graph representation of a simple MDP with
two states and two actions available in each state}{MDP-graph}

The optimal value of a state $s$, denoted by $V^*(s)$, represents the expected sum of discounted rewards
that an agent starting at state $s$ would obtain under the optimal policy. The optimal value of an
action $a$ taken from state $s$, denoted by $Q^*(s,a)$, represents the expected sum of discounted rewards 
that an agent would obtain after starting from state $s$, performing action $a$, and following the 
optimal policy thereafter. These values are connected with the equations:
\eq{Q^*(s,a) & = \sum_{s' \in S} T(s,a,s') \left[R(s,a,s') + \gamma V^*(s')\right] \\
    V^*(s) & = \max_{a \in A(s)}(Q^*(s,a))}\label {MDPequations}

The optimal values of the states therefore are the solutions to the set of equations:
$$\label{val_rec}
  V^*(s) = \max_{a \in A(s)} 
  \left(\sum_{s' \in S} T(s,a,s') 
  \left[R(s,a,s') + \gamma V^*(s')\right]\right)
$$

These equations can be solved using traditional dynamic programming 
techniques such as \textit{Value Iteration} \cite{bellman}.
The optimal policy, given the optimal values of the states, is:
$$
  \pi^*(s) = arg \max_a 
  \left(\sum_{s' \in S} T(s,a,s') 
  \left[R(s,a,s') + \gamma V^*(s')\right]\right)
$$

When such a computation is not feasible or desirable, \textit{Q-learning} \cite{watkins-art} is employed, where it performs a local incremental computation every time a decision is made. 

Q-learning maintains estimates of the Q-values (the values of the state-action pairs) and updates them
with every new experience using the equation:
$$Q(s,a) \leftarrow (1-\alpha) Q(s,a) + \alpha \left(r + \gamma \max_{a' \in A(s')} Q(s',a')\right)$$

The parameter $\alpha$ is known as the \textit{learning rate} and controls how fast the agent
learns from new experiences, $r$ is the reward obtained and $s'$ is the new observed state.

\textbf{Static vs adaptive state space partitioning.} 
This optimality is achieved under the assumption that the behavior
of the system is the same each time it finds itself in some specific state. This means
that the states need to be fine-grained enough to capture all the system complexity. Therefore, a complex system requires a large RL model. Nevertheless, an RL model of this scale is not only unrealistic to represent in memory, but even more so impossible to train, since the amount of experiences required to learn
the behavior of the system would also be exponential to the parameter number and grain. In such cases, methods that partition the state space adaptively can be employed \cite{chapman, qdt, mccallum1996reinforcement, continuousutree} in order to minimize the total number of states. Nevertheless, such methods need to carefully detect both the relevant parameters and their values to perform the partitioning.

\textbf{Model-based vs model-free.} 
There are two approaches through which an optimal policy can be calculated.

 In the \emph{model-based} approach, the agent attempts to find the exact behavior of the world,
in the form of the transition and reward functions, and then calculate the optimal policy
using dynamic programming approaches \cite{bellman}, or using alternative algorithms such as \textit{Prioritized
Sweeping} \cite{prioritized-sweeping} to decrease the required calculation in each step.

Oppositely, in the \emph{model-free} approach the agent evaluates the
effectiveness of the actions without learning the exact behavior of the world.
The most common model-free approach is the \textit{Q-learning} algorithm \cite{watkins-art}, which maintains estimates of Q-values (the values of the state-action pairs) and updates them with every new experience. 
One big advantage of Q-learning is that it is computationally and memory-wise efficient. 
Since it does not construct a full model of the world, it only needs to store the values of the Q-states. Also, each
update only involves calculating the maximum Q-value in the resulting state and performing a
simple update on the value of the initial state. 
%
However, Q-learning is limited to perform only local updates to the values. At each 
step, only the value of the performed action can be updated,
and as a result, only the policy of the initial state is updated. 
Additionally, in order to take into account both past and recent
experiences, Q-learning only partially updates the values of actions, 
which also slows down the learning process.
Therefore, Q-learning requires a larger amount of experiences to
converge to an optimal policy, especially if the agent performs a long
sequence of actions to obtain a reward \cite{rl-survey}.

Resource management decisions in a large computing environment are executed with a frequency of minutes at least. Therefore, collecting a very large training set of experiences would be unrealistic, and model-free approach is hard. Oppositely, the fact that a full-model approach requires a smaller training set in order to converge to an optimal policy, makes it a preferable choice. Furthermore, the advantage that Q-learning offers, which is
the extremely low requirements in memory and computational power, is not important for a cluster or cloud. For these reasons, we choose to adopt a full-model approach to solve the problem of resource management in a large elastic computing environment.

\begin{table}[t]
\centering
\caption{Terminology}
\vspace{-0.1in}
\label{terminology}
  \begin{adjustbox}{max width=0.5\textwidth}
\begin{tabular}
{|p{3cm}|p{1.5cm}|p{6.5cm}|}
\hline
\multicolumn{1}{|c|}{\textbf{Term}} & \textbf{Type} & \textbf{Semantics} \\ \hline
collect\_measurements & function & collects real system measurements \\ \hline
state & vector & stores state values with respect to values in collect\_measurements \\ \hline
select\_action & function & selects an action given the current state s \\ \hline
m, m' & vector & an ordered set of real parameter measurements \\ \hline
r & real & stores the value of a reward \\ \hline
get\_reward & function & calculates the reward given a measurement m \\ \hline
a & integer & stores the value of an action \\ \hline
e & vector & stores the measurements m, m' for states s, s' respectively, the action a and the reward r from s to s' \\ \hline
experiences & 2D vector & stores the experiences e for a pair of states s and s' \\ \hline
transitions & 3D vector & stores the number of transitions from state s to s' taking action a \\ \hline
rewards & 3D vector & stores the accumulated reward r for transitions from state s to s' taking action a \\ \hline
optimal\_action & function & returns the optimal action a for a state s \\ \hline
\end{tabular}
\end{adjustbox}
\end{table}

\section{The MDP\_DT Algorithm} \label{sec:implementation}
In this section we present our novel \eat{, open-source \cite{ommitedforblindreview2}} algorithm for elastic resource management based on adaptive state space partitioning and the use of decision trees.

\subsection{Overview}

\begin{algorithm}[t]
\scriptsize
\caption{MDP\_DT Algorithm}\label{alg:MDPDT}
\begin{algorithmic}[1]
\State $m = collect\_measurements()$
\While {$True$}
  \State $s = state(m)$
  \State $a = select\_action(s)$
  \State $execute\_action(a)$
  \State $sleep()$
  \State $m' = collect\_measurements()$
  \State $r = get\_reward(m')$
  \State $e = (m, m', a, r)$
  \State $UpdateMDPModel(e)$
  \State $UpdateModelValues(s)$
  \State $ApplySplittingCriterion(s)$
  \State $m = m'$
\EndWhile\\

\Procedure{UpdateMDPModel}{e}
  \State $m, m', a, r = e$
  \State $s = state(m)$
  \State $s' = state(m')$
  \State $experiences(s, s').add(e)$
  \State $transitions(s, a, s')$++
  \State $rewards(s, a, s')$ += $r$
\EndProcedure\\

\Procedure{UpdateModelValues}{s}
  \State \textbf{switch} {$update\_algorithm:$}
    \State \ \ \ \textbf{case} SINGLE\_UPDATE:
       \State \ \ \ \ \ \ $update\_q\_values(s)$
    \State \ \ \ \textbf{case} VALUE\_ITERATION:
       \State \ \ \ \ \ \ $value\_iteration()$
    \State \ \ \ \textbf{case} PRIORITIZED\_SWEEPING:
       \State \ \ \ \ \ \ $prioritized\_sweeping(s)$
\EndProcedure\\

\Procedure{ApplySplittingCriterion}{s}
  \State \textbf{switch} {$splitting\_criterion:$}
    \State \ \ \ \textbf{case} PARAMETER\_TEST:
       \State \ \ \ \ \ \ $split\_parameter\_test(s)$
    \State \ \ \ \textbf{case} Q\_VALUE\_TEST:
       \State \ \ \ \ \ \ $split\_q\_value\_test(s)$
\EndProcedure
\end{algorithmic}
\end{algorithm}

\textbf{Goal.} The computing environment consists of the \emph{system} and the \emph{workload}, and the \emph{system resources}, which are elastic and are used to accommodate the workload execution. The parameters of the system resources and the parameters of the workload are used to model the environment.  These can be multiple, behave in an interrelated or independent manner, play a significant or insignificant role in the performance of workload execution, and may be related to:

\textbf {Parameters of system resources:} cluster size, the amount of VMs, the amount of RAM per VM, the number of virtual CPUs (VCPUs) per VM, the storage capacity per VM, network characteristics, etc.

\textbf {Parameters of the workload:} CPU and network utilization, I/O reqs/sec, average job latency, etc.

Our goal is to accommodate the workload execution by adapting in a dynamic and online manner the system resources, so that the workload is executed efficiently and resources are not over-provisioned. We need to do this without knowing in advance the characteristics, role and interaction of the parameters of both system resources and workload.

\textit{Motivating Example}: A start-up company hosts services in a public IaaS cloud. The company employs a distributed data-store (for instance, a NoSQL database) to handle user-generated workload that consists, e.g., of a mix of read and write requests. The administrator wants to optimize a given business policy, typically of the form ``maximize performance while minimizing the cost of operating the infrastructure'' under variable and unpredictable loads. For example, if the company offers user-facing low-latency services, the performance can be measured based on throughput, and the cost can be the cost of renting the underlying IaaS services.
In order to achieve this she wants to use an automated mechanism to perform one or both of the following: (i) scale the system, (e.g., change the cluster size, the RAM size, etc.) or (ii) re-configure the system (e.g., increase cache size, change replication factor, etc). Such a mechanism may implement a rule-based technique, like one of the aforementioned frameworks \cite{aws_autoscale, azure_autoscale, softlayer_autoscale, rackspace_autoscale, google_autoscaling}, which monitors the value of representative performance parameters, such as CPU, RAM usage, incoming workload, etc. and perform specific scaling and reconfiguration actions. Such an approach has the following shortcomings. First, it is difficult to detect which, among the numerous, system parameters affect performance, as they are application and workload dependent. For example, in a specific write-heavy scenario the CPU usage may not be affected, as the bottleneck is mainly due to I/O operations and, thus, a CPU based rule will not work. Second, even if we can detect the parameters, it is difficult to determine which the respective actions should be. For example, even if we can conclude that for a write-heavy scenario we need a I/O rule, the appropriate action may not be obvious: for instance, increasing the RAM and cache size of existing servers to avoid I/O thrashing may be a better action than adding more servers. Third, it is difficult to detect the threshold values based on which actions need to be triggered. For example, a pair of thresholds on ``high''/``low'' CPU usage, a threshold on the number I/O ops or on memory usage are very application-specific and need a lot of fine tuning. For all three reasons, the translation of higher level business policies for the maximization of performance and minimization of cost into a rule-based approach that automatically scales and reconfigures the system is very difficult and error-prone.

\textbf{Solution.} We create a model of the computing environment by representing each selected parameter of system resources and workload with a distinct dimension. Therefore the environment is modelled by a multi-dimensional space in which all possible states of the environment can be represented, with variable detail. We create a novel MDP algorithm that starts with one or a few model states that cover the entire multi-dimensional state space. The algorithm gradually partitions the coarse state space into finer states depending on observed measurements of the modelled environment parameters. The algorithm employs a decision tree in order to perform this dynamic and adaptive partitioning. At each state the algorithm takes an action in order to make transition toward another state. Such actions may change the values of some of the parameters of the system resources, e.g. change of (i) the size of the cluster in terms of machines and (ii) the number of VMs. Nevertheless, even though the actions change only some parameters of the system resources, they may affect many more such parameters, as well as parameters of the workload. The type of actions allowed is given as an input to the algorithm.

\textit{Motivating Example - continued}: Employing our MDP approach the administrator of the startup company needs only to provide the following: a) a list of parameters she considers important to performance (even if some of them may turn out to not affect performance, at least for the observed states), b) a list of available scaling or reconfiguration actions, c) a high-level user-defined policy in the form of a reward function that encapsulates the maximization of performance and minimization of cost, and, optionally d) some initial knowledge in the form of a ``training set'' to speed up the learning process (in Section \ref{sec:simulations} we give a concrete example of parameters, actions and policies). Then, \emph{MDP\_DT} algorithm adaptively detects both the set of parameters that affect the reward and the appropriate scaling and reconfiguration actions that maximize the reward.


\textbf{Description of the algorithm.}
The \emph{MDP\_DT} algorithm is presented in Alg. \ref{alg:MDPDT} and Table \ref{terminology} summarizes its terminology.
\emph{MDP\_DT} starts with a single tree node (the root of the decision tree), which corresponds to one state covering the entire state space of the model of the environment. A vector \emph{state} is maintained for all possible states of the environment. Each element $s$ in \emph{state} corresponds to a list of Q-states, holding the number of transitions \emph{transitions} and the sum of rewards \emph{rewards} towards each state $s'$ in the model, along with the total number of times the action has been taken. From this information we can easily calculate the transition and reward functions. The current state $s$ is represented by a set of measurements $m$ that contain the names  and current values for all the parameters of the environment. The state $s'$ to which action $a$ leads is represented by a respective set of measurements $m'$. Given the current state $s$ (and corresponding measurements $m$), the algorithm selects an action $a$, the action is performed, and the algorithm collects the measurements $m'$ for the new state $s'$ of the environment, for which is calculates the reward $r$. This transition experience $e = (m,a,m',r)$ is used to update the MDP model, the model values and split the state $s$, using procedures \emph{UpdateMDPModel},  \emph{UpdateMDPValues} and \emph{ApplySplittingCriterion}, respectively. 
Procedure  \emph{UpdateMDPModel} saves the experience $e = (m, m', a, r)$ in the $experiences$ vector in the place corresponding to the pair of $s, s'$, increases the number of transitions for the pair of $s,s'$ and adds the new reward $r$ to the accumulated reward for the pair of $s, s'$.
Procedure  \emph{UpdateMDPValues} updates the Q-state values for state $s$ by employing one of the classical update algorithms: single update (i.e. Equation 1), value iteration, and prioritized sweeping.  
Procedure \emph{ApplySplittingCriterion} considers splitting state $s$ into two new states, based on a splitting test. For this we propose two splitting criteria, the \emph{parameter test} and the \emph{Q-value test}, which we present in the following.

\begin{algorithm}[t]
\scriptsize
\caption{Parameter Test}\label{alg:parameter_test}
\begin{algorithmic}[1]
\Procedure{SplitParameterTest}{s}
\State $a = optimal\_action(s)$
\State $es = \{e\ |\ e \in experiences(s, *), e.a = a\}$
\State $e_+ = \{e\ |\ e \in es, q\_value(e) \ge value(a)\}$
\State $e_- = \{e\ |\ e \in es, q\_value(e) < value(a)\}$
\State $lowest\_error = 1$
\For{$p$ \textbf{in} $parameters$}
  \State $p_- = \{e.m[p]\ |\ e \in e_-\}$
  \State $p_+ = \{e.m[p]\ |\ e \in e_+\}$
  \State $error\_prob = stat\_test(p_-, p_+)$
  \If {$error\_prob < lowest\_error$}
    \State $lowest\_error = error\_prob$
    \State $best\_p = p$
    \State $best\_p_- = p_-$
    \State $best\_p_+ = p_+$
  \EndIf
\EndFor
\If {$lowest\_error \le max\_type\_I\_error$}
  \State $mean_- = mean(best\_p_-)$
  \State $mean_+ = mean(best\_p_+)$
  \State $split\_point = (mean_- + mean_+) / 2$
  \State $split(s, best\_p, split\_point)$
\EndIf
\EndProcedure
\end{algorithmic}
\end{algorithm}

\begin{algorithm}[t]
\scriptsize
\caption{Q Value Test}\label{alg:q_value_test}
\begin{algorithmic}[1]
\Procedure{SplitQValueTest}{s}
\State $a = optimal\_action(s)$
\State $es = \{e\ |\ e \in experiences(s, *), e.a = a\}$
\State $N = length(es)$
\State $lowest\_error = 1$
\For {$p$ \textbf{in} $parameters$}
  \State $sort\_by\_param(es, p)$
  \For {$i$ \textbf{in} $1..N-1$}
    \If {$es[i].m[p] = es[i+1].m[p]$}
      \State \textbf{continue}
    \EndIf
    \State $q_- = \{q\_value(e)\ |\ e \in es[1..i]\}$
    \State $q_+ = \{q\_value(e)\ |\ e \in es[i+1..N]\}$
    \State $error\_prob = stat\_test(q_-, q_+)$
    \If {$error\_prob < lowest\_error$}
      \State $lowest\_error = error\_prob$
      \State $best\_p = p$
      \State $split\_point = (e.m[p] + e'.m[p]) / 2$
    \EndIf
  \EndFor
\EndFor
\If {$lowest\_error \le max\_type\_I\_error$}
  \State $split(s, best\_p, split\_point)$
\EndIf
\EndProcedure\\

\Function{QValue}{e}
\State $m, m', a, r = e$
\State $s' = state(m')$
\State $a' = optimal\_action(s')$
\State \Return $r + \gamma \cdot value(a')$
\EndFunction
\end{algorithmic}
\end{algorithm}

\subsection{Splitting Criteria} \label{split-crit}

The proposed splitting criteria, \emph{parameter test} and the \emph{Q-value test}, have two strengths. First, the Q-value derived from each experience is calculated using the current, most accurate values of the states instead of the values at the time the action was performed. 
    Second, the partitioning of the experiences is done by comparing them to the current value of state $s$ instead of partitioning them to experiences that increased or decreased the Q-value at the time of their execution. These two features allow the reliable use and re-use of experiences collected early in the training processing, at which point the values of the states were not yet known, throughout the lifetime and adaptation of the model.
    
    \subsubsection{Parameter test}
    Procedure \emph{SplitParameterTest} presented in Alg. \ref{alg:parameter_test} implements the splitting criterion \emph{parameter test} which works as follows.
     From the experiences $e = (m,a,m',r)$ stored in the \emph{experiences} vector for every pair of $s, s'$, we isolate the experiences where the action $a$ was the optimal
    action for state $s$ (i.e., $a$ led to the highest Q-value). For each of these experiences,
    we find the state $s'$ in the current model that corresponds to $m'$ using the decision tree,
    and calculate the value $q(m,a) = r + \gamma V(s')$.
    We then partition this subset of experiences to two lists $e_-$ and $e_+$ by comparing $q(m,a)$ with
    the current value of the optimal action for state $s$.

    For each parameter $p$ we divide the values of $p$ for the measurements
    in $e_-$ and $e_+$ in two lists $p_-$ and $p_+$, and run a statistical test on $p_-$
    and $p_+$ to determine the probability that the two samples come from the
    same population. We choose to split the parameter with the lowest such probability, 
    as long as it is lower than the error $max\_type\_I\_error$, else the procedure aborts. If the 
    split proceeds, the splitting point is the average of the means of $p_-$ and $p_+$.


  \eat{  The strengths of this splitting criterion are two. First, the Q-value derived from each experience is calculated using the 	current and, therefore, most accurate values of the states instead of the values at the time the action was performed. 
    Second, the partitioning of the experiences is done by comparing them to the current value of state $s$ instead of partitioning them to experiences that increased or decreased the Q-value at the time of their execution. These two features allow the reliable use and re-use of experiences collected early in the training processing, at which point the values of the states were not yet known, throughout the lifetime and adaptation of the model.
    }

\begin{figure*}
    \centering
    \begin{subfigure}[b]{0.23\textwidth}
        \includegraphics[width=\textwidth]{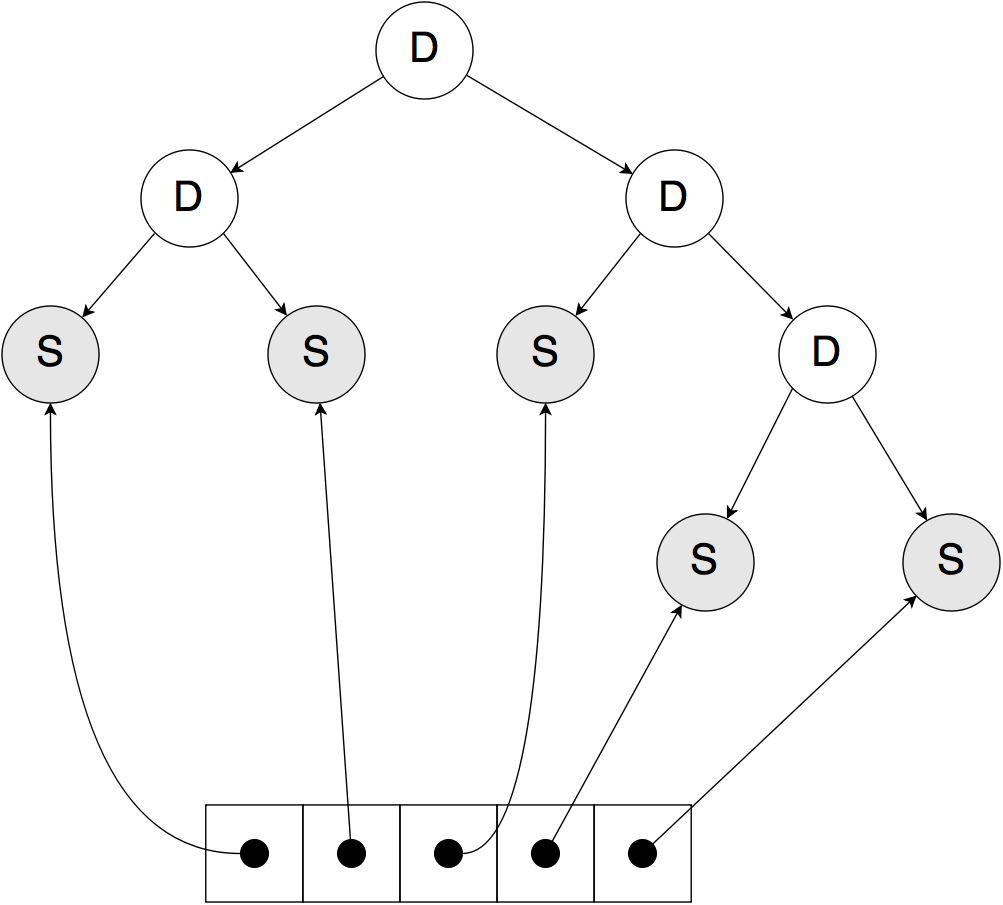}
        \caption{The decision tree before splitting a state \\ \ \\ \ }
        \label{fig:gull}
    \end{subfigure}
    ~
    \begin{subfigure}[b]{0.23\textwidth}
        \includegraphics[width=\textwidth]{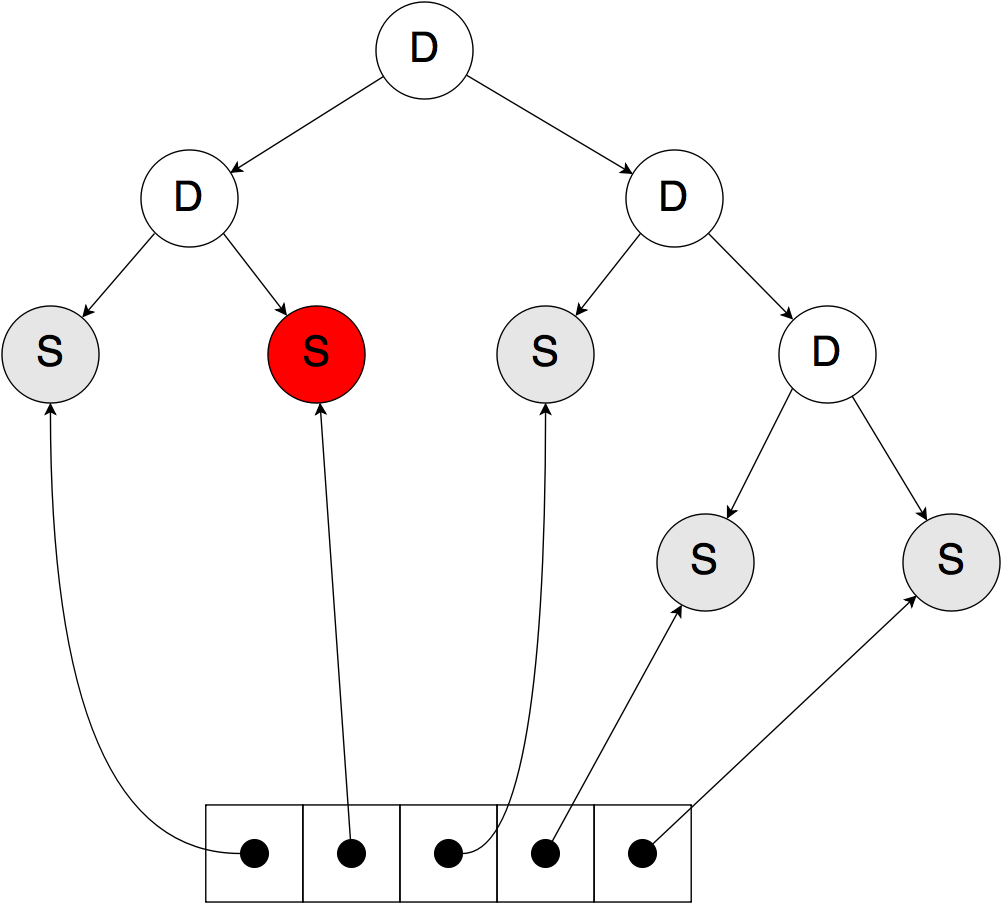}
        \caption{ Experiences concering the state to be split are transferred to temporary storage, respective transition and reward information is removed.}
        \label{fig:tiger}
    \end{subfigure}
    ~
    \begin{subfigure}[b]{0.23\textwidth}
        \includegraphics[width=\textwidth]{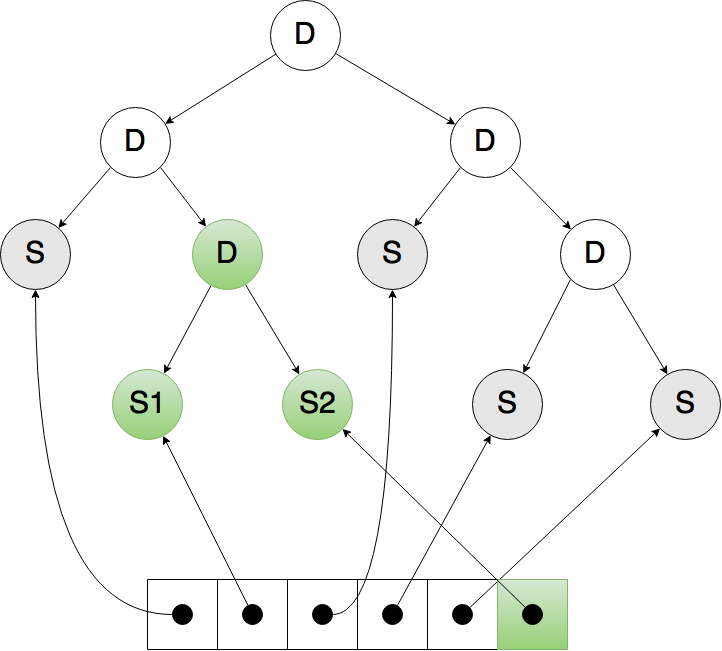}
        \caption{The state is replaced by a new decision node and two children corresponding to the two new states. \\ \ }
        \label{fig:mouse}
    \end{subfigure}
    ~
    \begin{subfigure}[b]{0.23\textwidth}
        \includegraphics[width=\textwidth]{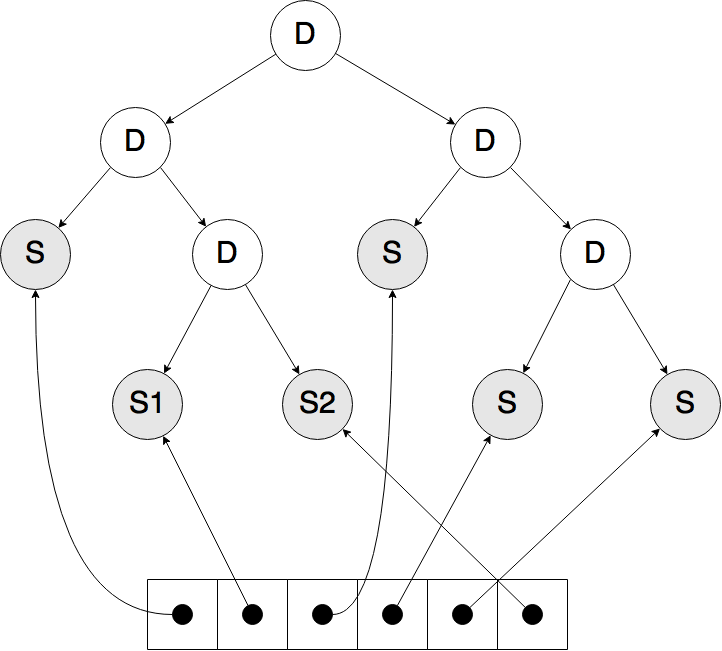}
        \caption{Experiences in temporary storage are used to train the new states, and calculate the new transitions and rewards.}
        \label{fig:bat}
    \end{subfigure}
    \vspace{-0.05in}
    \caption{Splitting a state $s$ into two new states $s_1, s_2$: $s_1$ replaces $s$ in the $state$ vector and the
             $s_2$ is appended at the end.}\label{fig:mdpdt}
             \vspace{-0.15in}
\end{figure*}

    \subsubsection{Q-value test}
    Procedure \emph{SplitQValueTest} presented in Alg.\ref{alg:q_value_test} implements the splitting criterion \emph{Q-value test} which works as follows.
    Again, from the experiences $e = (m,a,m',r )$ related to pairs of $s, s'$, we isolate the experiences where the action $a$ is the current optimal action for $s$. For each such experience, we find the state $s'$ that
    corresponds to $m'$ using the current decision tree and calculate $q(m,a) = r + \gamma V(s')$. 

    For each parameter $p$ of the system, we sort these experiences based on the value of $p$,
    and consider splitting in the midpoint between each two consecutive unequal values. For that
    purpose, we run a statistical test on the Q-values in the two resulting sets of experiences, 
    and choose the splitting point that produces the lowest probability that represents the fact the two sets of 
    values are statistically indifferent, as long as that probability is less than the error $max\_type\_I\_error$. 
%
 This criterion performs a straightforward comparison of subsets of experiences with respect to the optimality of the taken action. It is a criterion \emph{that we adapted from the \textit{Continuous U Tree} algorithm \cite{continuousutree}, 
    and} resembles splitting criteria used in traditional algorithms for decision tree induction such as C4.5 \cite{quinlan1993c4}. 
%
    Additionally, we experimented with splitting in the midpoint between
    the two unequal consecutive measurements that are closest to the median. This considers a single splitting point per parameter, which splits the experiences approximately equally in the two new states.

\begin{algorithm}[t]
\scriptsize
\caption{Splitting a State}\label{alg:splitting_a_state}
\begin{algorithmic}[1]
\Procedure{Split}{s, param, point}
\State $transitions(*,*,s) = 0$
\State $rewards(*,*,s) = 0$
\State $es = \{e\ |\ e \in experiences(s,*) \cup experiences(*,s)\}$
\State $experiences(s,*) = experiences(*,s) = []$
\State $replace\_with\_decision\_node(s, param, point)$
\For {$e$ \textbf{in} $es$}
   \State $UpdateMDPModel(e)$
\EndFor
\State $UpdateModelValues(s)$
\EndProcedure
\end{algorithmic}
\end{algorithm}
   
\subsection{Performing Splits}

Once a split has been decided by the \emph{parameter test} or the \emph{Q-value test} for a state $s$, the splitting is performed by procedure \emph{Split} shown in Alg. \ref{alg:splitting_a_state}. Figure \ref{fig:mdpdt} showcases this procedure.
For state $s$, all transition and reward information is removed from the $transitions$ and $rewards$ vectors, respectively. Also, the experiences $e$ that involve $s$ as the starting or the ending state are accumulated and stored in temporary storage $es$. These experiences are removed from the $experiences$ vector, because state $s$ will be substituted by two new states $s_1, s_2$. This is performed in two steps: First, the node in the decision tree that corresponds to state $s$ is replaced with a new decision node, and two children nodes, (leafs), that correspond to the two new states that result from splitting $s$. Second, one of the new states takes the place of the split state $s$ in the $state$ vector and the second new state takes a new position appended at the end of the $state$ vector. The leaf nodes are linked to the positions in the $state$ vector of the corresponding states. The $reward$, $transition$ and $experiences$ vectors are updated and extended accordingly, with new elements for combinations of $s_1, s_2$ with all other states.
The obsolete experiences $es$ are used to retrain the new states: For each $e = (m, a, m', r)$, the new states $s$ and $s'$ are found using the updated decision tree, and the respective positions in the $reward$ and $transition$ vectors are updated.

\subsection{Statistical Tests}\label{app:tests}

The two splitting criteria include a statistical test to determine whether the two groups of compared values are statistically different from each other. For that purpose, we employ four different statistical tests.

\textbf{Student's t-test.} 
 The statistic for this test is calculated using the formula:
      \eq{t = \frac{\overline{X}_1 - \overline{X}_2}
          {s_{X_1X_2} \cdot \sqrt{\frac{1}{n_1} + \frac{1}{n_2}}}}
      where $s_{X_1X_2}$ is an estimator of the common standard deviation of the two samples
      given by:
      \eq{s_{X_1X_2} = \sqrt{\frac{(n_1-1)s_{X_1} + (n_2-1)s_{X_2}}{n_1+n_2-2}}}
      The quantity $n_1+n_2-2$ is the total number of degrees of freedom. This test was also
      used by \cite{qdt}, and is a very common way to test the similarity of two samples.

\textbf{Welch's test. } 
	 The statistic in this case is given by:
      \eq{t = \frac{\overline{X}_1 - \overline{X}_2}{\sqrt{\frac{s_1^2}{n_1} + \frac{s_2^2}{n_2}}}}
      where $s_1$ and $s_2$ are the unbiased estimators of the variance of the two samples.
      The degrees of freedom for this test are given by the \textit{Welch-Satterthwaite equation}:
      \eq{v = \frac{\left(\frac{s_1^2}{N_1} + \frac{s_2^2}{N_2}\right)^2}
                   {\frac{s_1^4}{N_1^2v_1} + \frac{s_2^4}{N_2^2v_2}}}
      This test has been proposed by \cite{welch1, welch2} as the default way to test the
      statistical similarity of two samples when the equality of the variances is not known 
      beforehand, over the \textit{Student's t-test}.

\textbf {Mann Whitney U test.}
      In this test, if the compared samples are small, it can easily be calculated by making every possible
      comparison between the elements of the two groups, and counting the amount of times
      the elements of each group wins (giving 0.5 to each group for ties). For larger samples
      the statistic can be calculated by ranking all the elements of the two groups in
      increasing order based on their value, adjusting the ranks in case of ties to the
      midpoint of unadjusted rankings, and summing up the ranks in the two groups \cite{wiki-mann}.
      The U statistic is then given by:
      \eq{U_1 = R_1 - \frac{n_1(n_1+1)}{2},\ U_2 = R_2 - \frac{n_2(n_2+1)}{2}}
      where $R_1$ and $R_2$ are the sums of ranks for the samples 1 and 2. The minimum value
      among $U_1$ and $U_2$ is then used to consult a significance table.

\textbf{Kolmogorov-Smirnov test.}
	  This test involves the calculation of the Kolmogorov Smirnov statistic:
      \eq{D_{n,n'} = sup_x|F_{1,n}(x) - F_{2,n'}(x)|}
      where $F_{1,n}$ and $F_{2,n'}$ are the \textit{empirical distribution functions} of
      the two samples, and $sup$ is the \textit{supremum function} \cite{wiki-kolmo}.
      The \textit{null hypothesis} is rejected at level $\alpha$ if:
      \eq{D_{n,n'} > c(\alpha)\sqrt{\frac{n+n'}{nn'}}}
      where $c(\alpha)$ is the inverse Kolmogorov distribution at $\alpha$. This was
      used for performing splits in \cite{wiki-kolmo}.

  \subsection{Splitting Strategy}\label{sec:splitstrategy}

By default, the \emph{MDP\_DT} algorithm attempts to split the starting state of each experience after this has been acquired, \emph{and} depending only on this. However, the effectiveness of the algorithm may be better if the splitting is performed after the acquisition of more than one experiences \emph{and/or} independently of these specific experiences. 
We investigate the above options with three basic \emph{splitting strategies}:
\begin{myitemize}
\item \textbf{Chain Split}: The goal of the \emph{Chain Split} strategy is to accelerate the division of the state space into finer states, by accelerating the growth of the decision tree. It attempts to split every node of the tree, regardless of whether it was involved in the current experience.  Then, if a split has been performed, it tries to split every node in the new tree once more. In the same way it re-tries to split every node in the new tree, until no more splits are performed. The rationale of this strategy is that the change in the value of the current state may affect also the value of other states, and, therefore, it should trigger the splitting not only of the current state, but also of others. 
\item \textbf{Reset Split}: The goal of the \emph{Reset Split} strategy is to try to correct splitting mistakes, by resetting the decision tree periodically, and by taking more accurate decisions after each reset, by taking into account all accumulated experiences.

\item \textbf{Two-phase Split:} The \emph{Two-phase Split} strategy splits based on accumulated experiences, rather than split after each new experience. \emph{Two-phase Split} performs the splitting on the existing decision tree periodically. 
Therefore, in this case the \emph{MDP\_DT} algorithm is versioned so that it has two phases, a \textit{Data Gathering} phase that collects data but does not perform any splits, and a \textit{Processing Phase} that the tree nodes are tested one by one to check if a split is needed, and if so, perform the splits. 
\end{myitemize}

In Section \ref{sec:simulation_splitstrategy} we experiment with versions  and combinations of the strategies.




\section{Simulation Results} \label{sec:simulations}

In this section we show experimental results on a simulation of the elastic computing environment. We experiment with a number of options that affect the performance of the \emph{MDP\_DT}
algorithm. The agent makes elasticity decisions that resize a cluster running a database under a varying incoming load. The load consists of read and write requests, and the cluster capacity (i.e., the maximum achievable query throughput) depends on its size as well as the percentage of the incoming requests that are reads. 
Specifically:

\begin{myitemize}
\item The cluster size can vary between 1 and 20 VMs.
\item The available actions are: the increase cluster size by one,
decrease the cluster size by one, or do nothing.
\item The incoming load is a sinusoidal function of time: 
      $load(t) = 50 + 50 sin\left(\frac{2\pi t}{250}\right)$.
\item The percentage of incoming read requests is a sinusoidal function of time with
a different period: $r(t) = 0.75 + 0.25 sin\left(\frac{2\pi t}{340}\right)$.
\item If $vms(t)$ is the number of VMs currently, the cluster capacity is: $capacity(t) = 10 \cdot vms(t) \cdot r(t)$.
\item The reward for each action depends on the state of the cluster after executing the action
and is given by $R_t = min(capacity(t+1), load(t+1)) - 3 \cdot vms(t+1)$.
\end{myitemize}

The reward function encourages the agent to increase the size of the cluster to the point
where it can fully serve the incoming load, but punishes it for going further than that. In order
for the agent to behave optimally, it needs to not only identify the way its actions affect the
cluster's capacity and the dependence on the level of the incoming load, but also the dependence
on the types of the incoming requests. 

In order to test the algorithm's ability to partition the state space in a meaningful manner, 
apart from the three relevant parameters (size of the cluster, incoming load and percentage of 
reads) the dimensions of the model include seven additional parameters, the values of which vary in a random manner. Four of 
them follow a uniform distribution within $[0,1]$, while the rest take integer values within
$[0,9]$ with equal probability. To be successful the algorithm needs to partition
the state space using the three relevant parameters and ignore the rest.

All experiments include a training phase and an evaluation phase. During the training phase, 
the selected action at each step is a random action with probability $e$, or the optimal
action with probability $1-e$ ($e$-greedy strategy). During the evaluation phase only optimal
actions are selected, as proposed by the algorithm. The metric according to which different options
are compared is the sum of rewards that the agent managed to accumulate during the evaluation phase.

\subsection{Statistical Significance} \label{sec:statistical-significance}

\begin{figure*}[ht!]
    \centering
    \begin{subfigure}[b]{0.32\textwidth}
        \includegraphics[width=\textwidth]{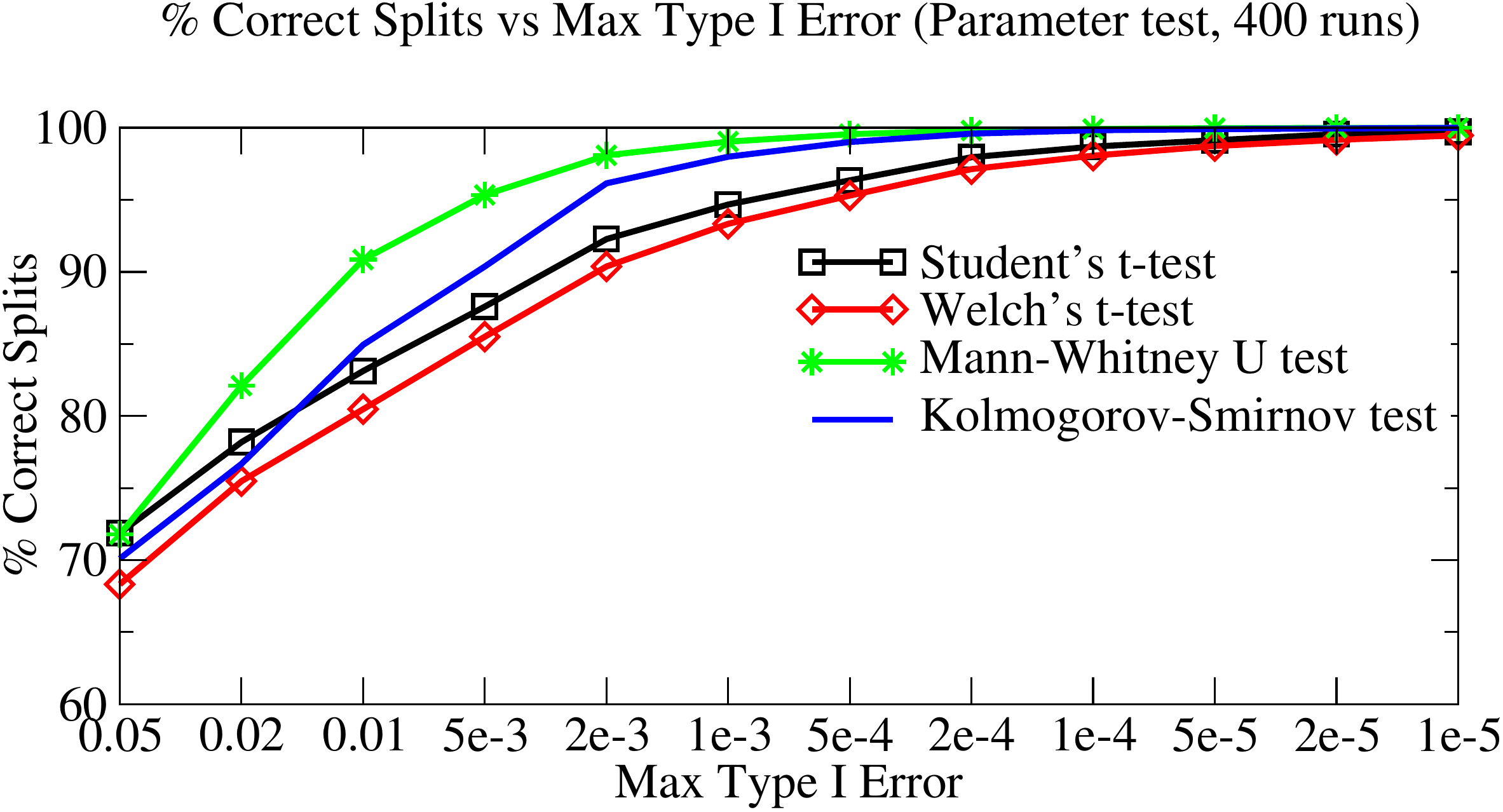}
        \caption{Parameter test}
    \end{subfigure}
    ~
    \begin{subfigure}[b]{0.32\textwidth}
        \includegraphics[width=\textwidth]{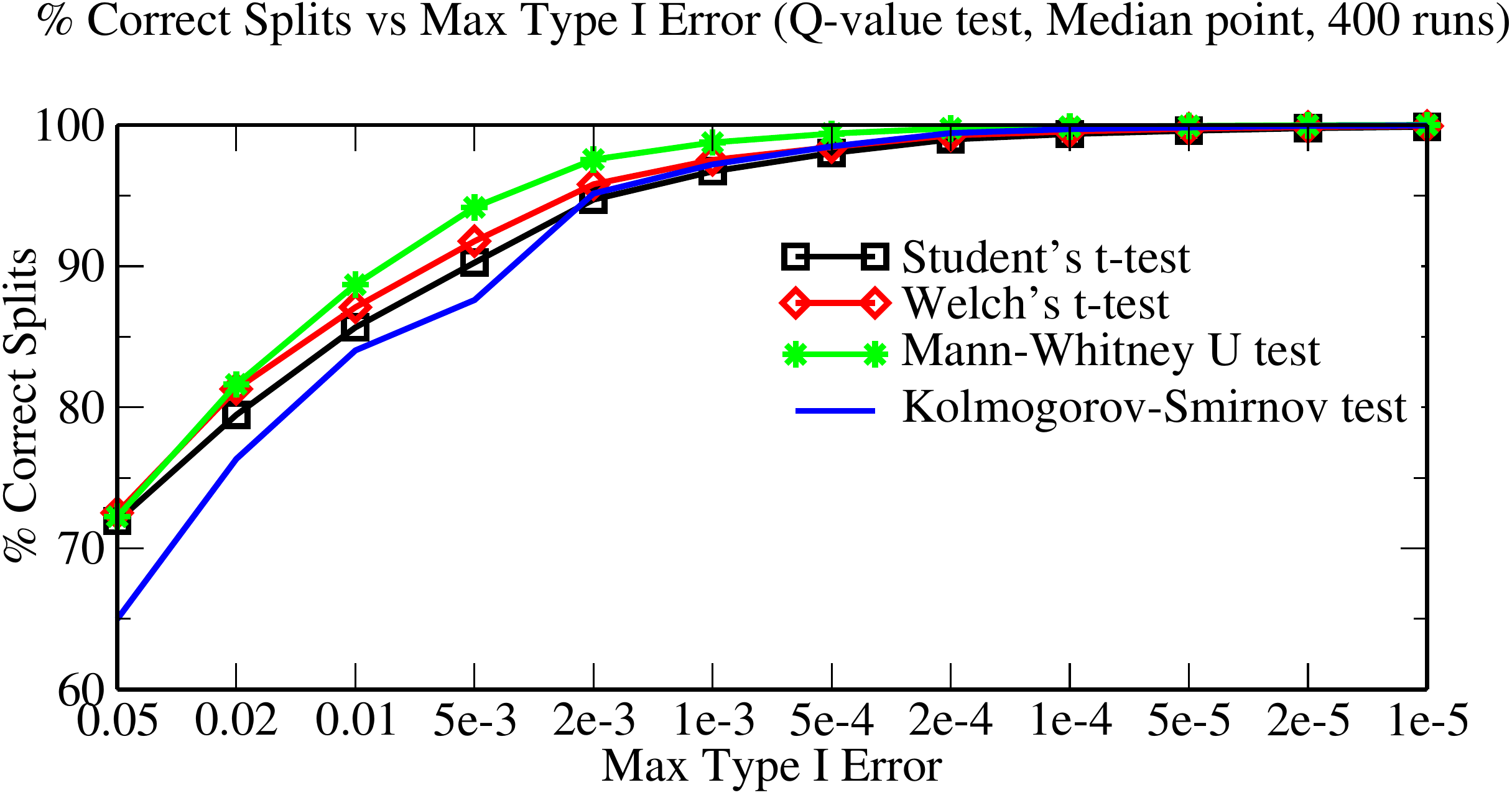}
        \caption{Q-value test, Median splitting point}
    \end{subfigure}
    ~
    \begin{subfigure}[b]{0.32\textwidth}
        \includegraphics[width=\textwidth]{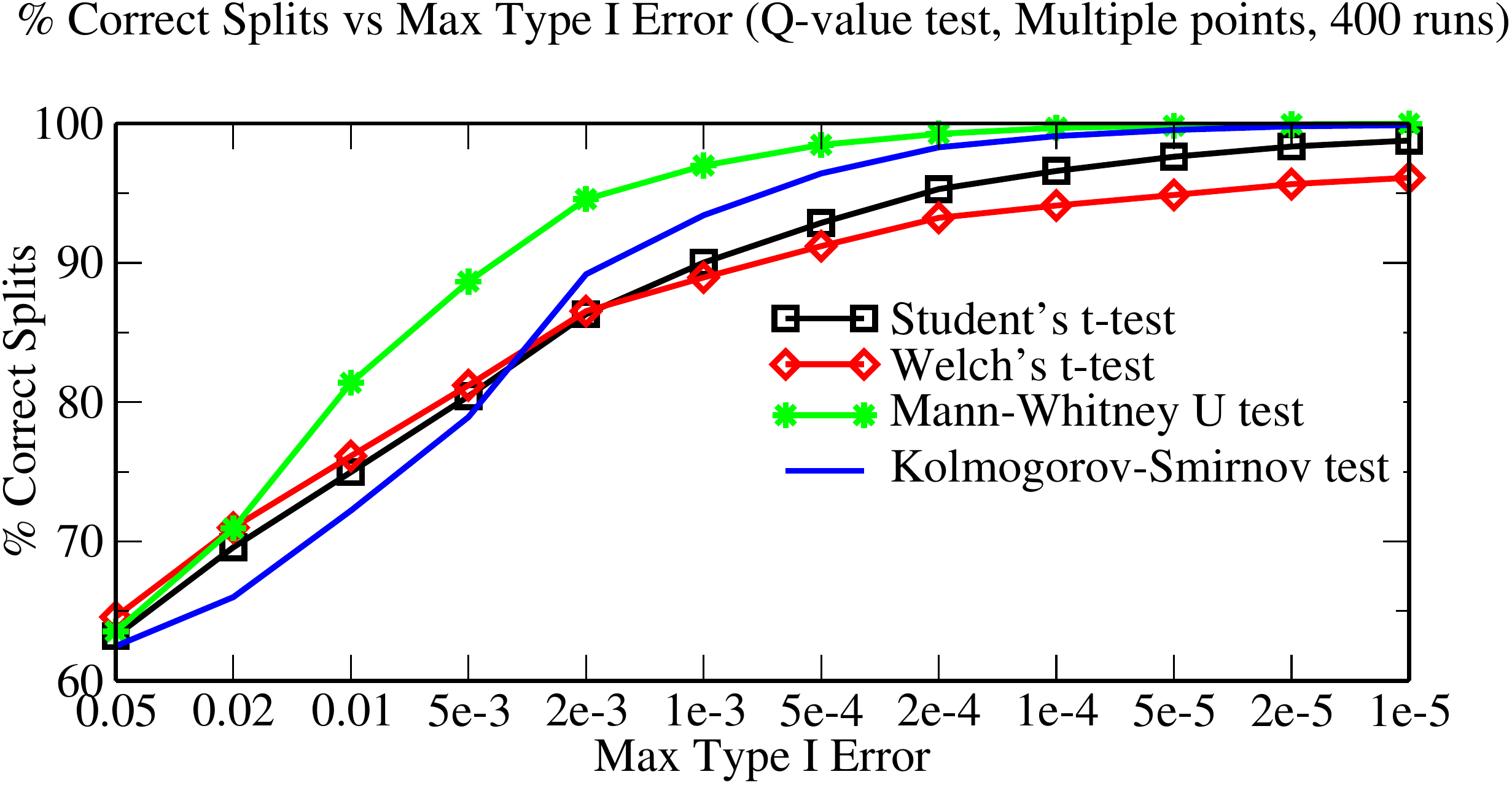}
        \caption{Q-value test, Multiple splitting points}
    \end{subfigure}
    \caption{Accuracy of the four statistical tests for three different splitting criteria as a
             function of the maximum error}
    \label{fig:stat-accuracy}
\end{figure*}

\begin{figure*}[ht!]
    \centering
    \begin{subfigure}[b]{0.32\textwidth}
        \includegraphics[width=\textwidth]{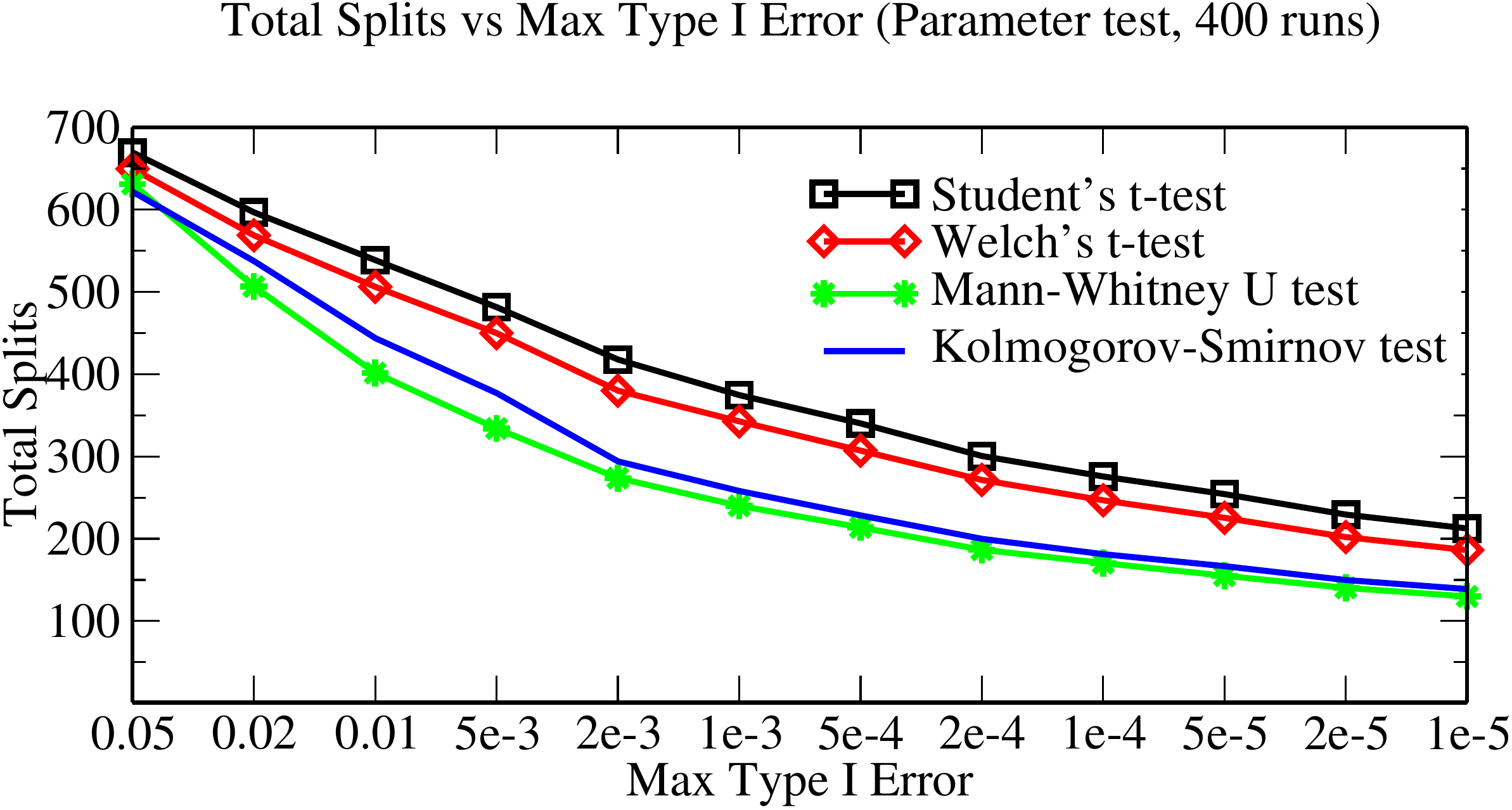}
		\vspace{-0.2in} 
        \caption{Parameter test}
    \end{subfigure}
    ~
    \begin{subfigure}[b]{0.32\textwidth}
        \includegraphics[width=\textwidth]{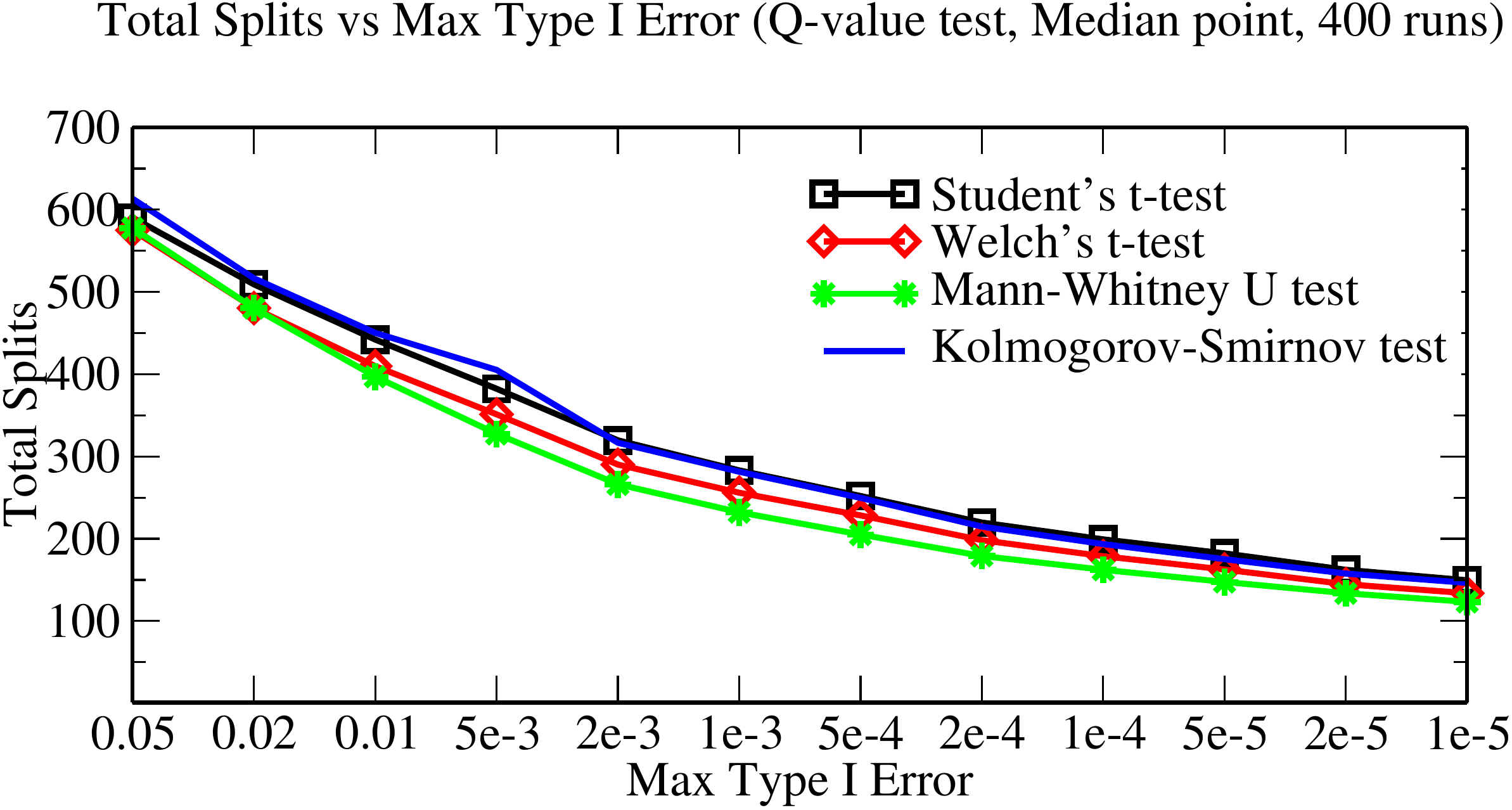}
		\vspace{-0.2in} 
        \caption{Q-value test, Median splitting point}
    \end{subfigure}
    ~
    \begin{subfigure}[b]{0.32\textwidth}
        \includegraphics[width=\textwidth]{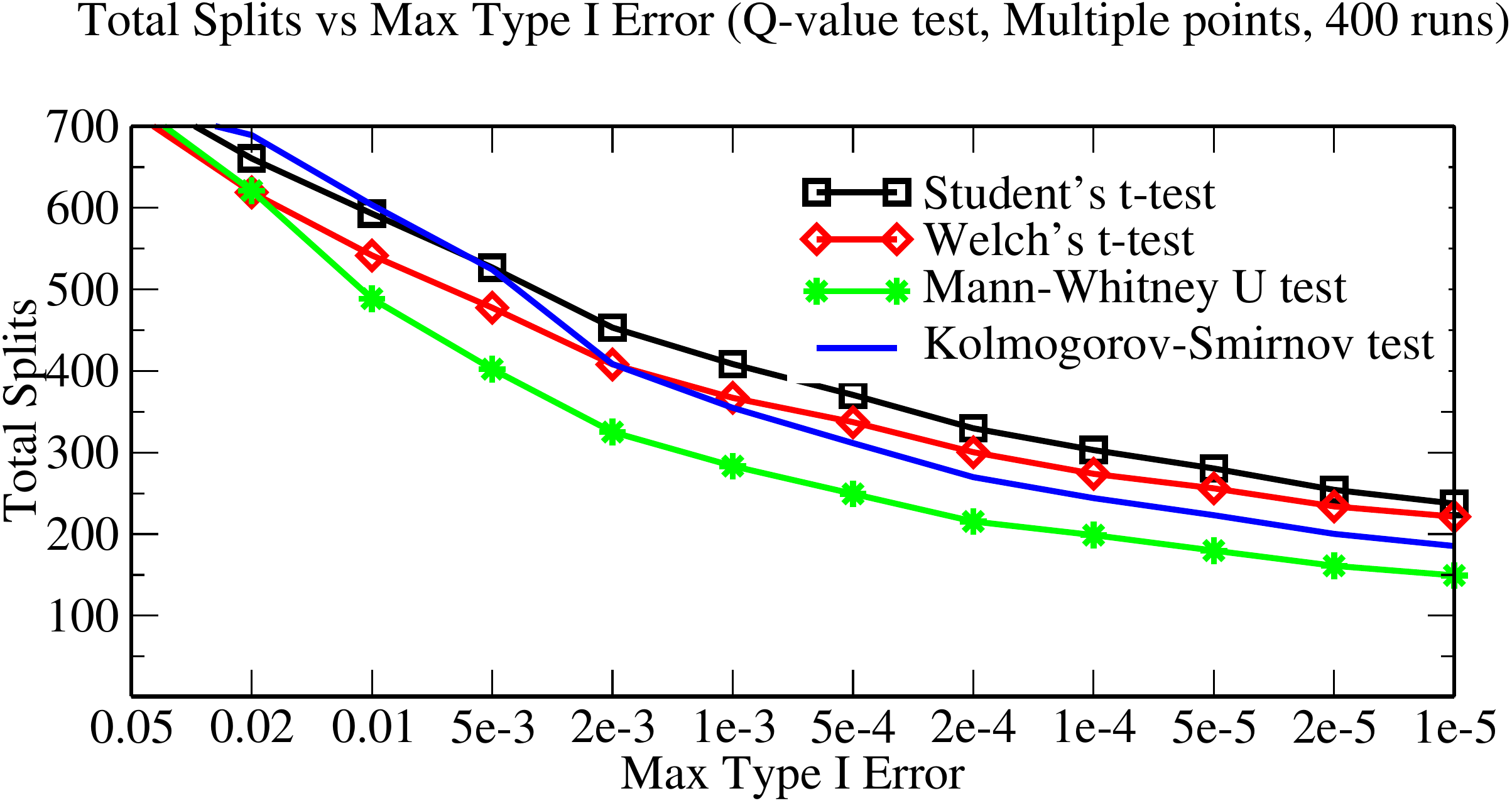}
		\vspace{-0.2in} 
        \caption{Q-value test, Multiple splitting points}
    \end{subfigure}
    \caption{Number of splits for the four statistical tests and three splitting criteria, as a
             function of the maximum error}
    \label{fig:stat-splits}
\end{figure*}

\begin{figure*}[ht!]
    \centering
    \begin{subfigure}[b]{0.32\textwidth}
        \includegraphics[width=\textwidth]{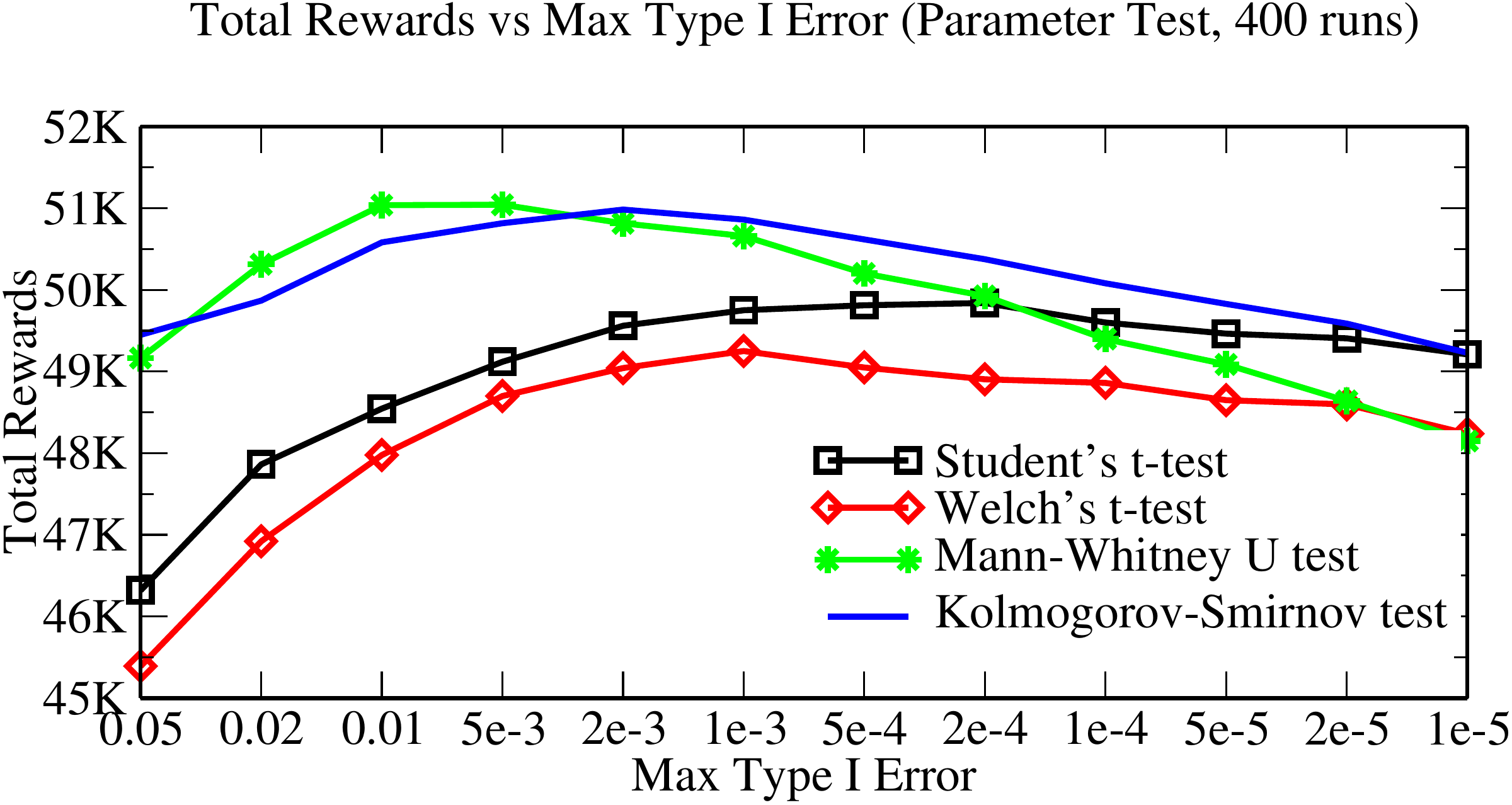}
		\vspace{-0.2in} 
        \caption{Parameter test}
    \end{subfigure}
    ~
    \begin{subfigure}[b]{0.32\textwidth}
        \includegraphics[width=\textwidth]{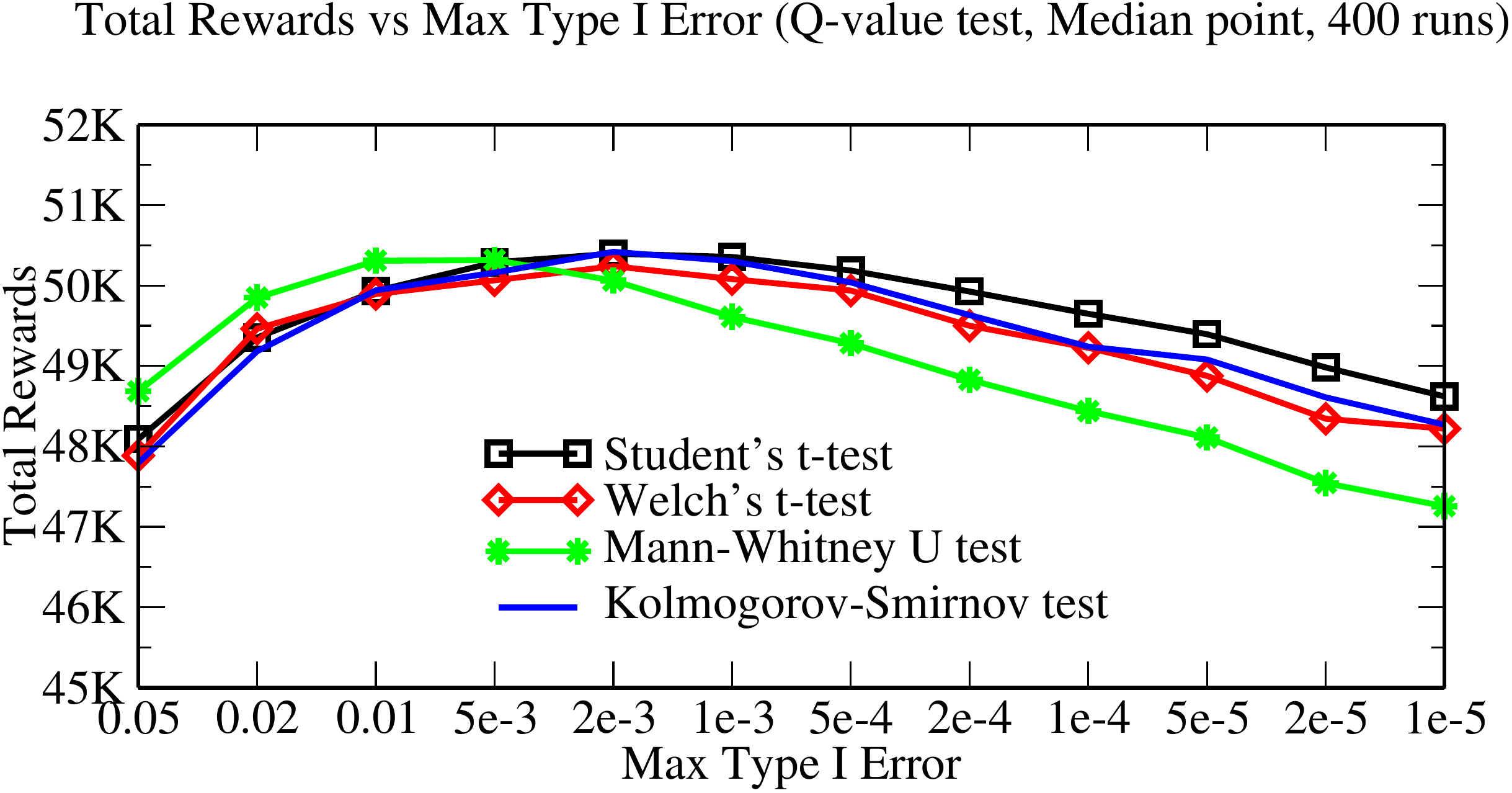}
		\vspace{-0.2in} 
        \caption{Q-value test, Median splitting point}
    \end{subfigure}
    ~
    \begin{subfigure}[b]{0.32\textwidth}
        \includegraphics[width=\textwidth]{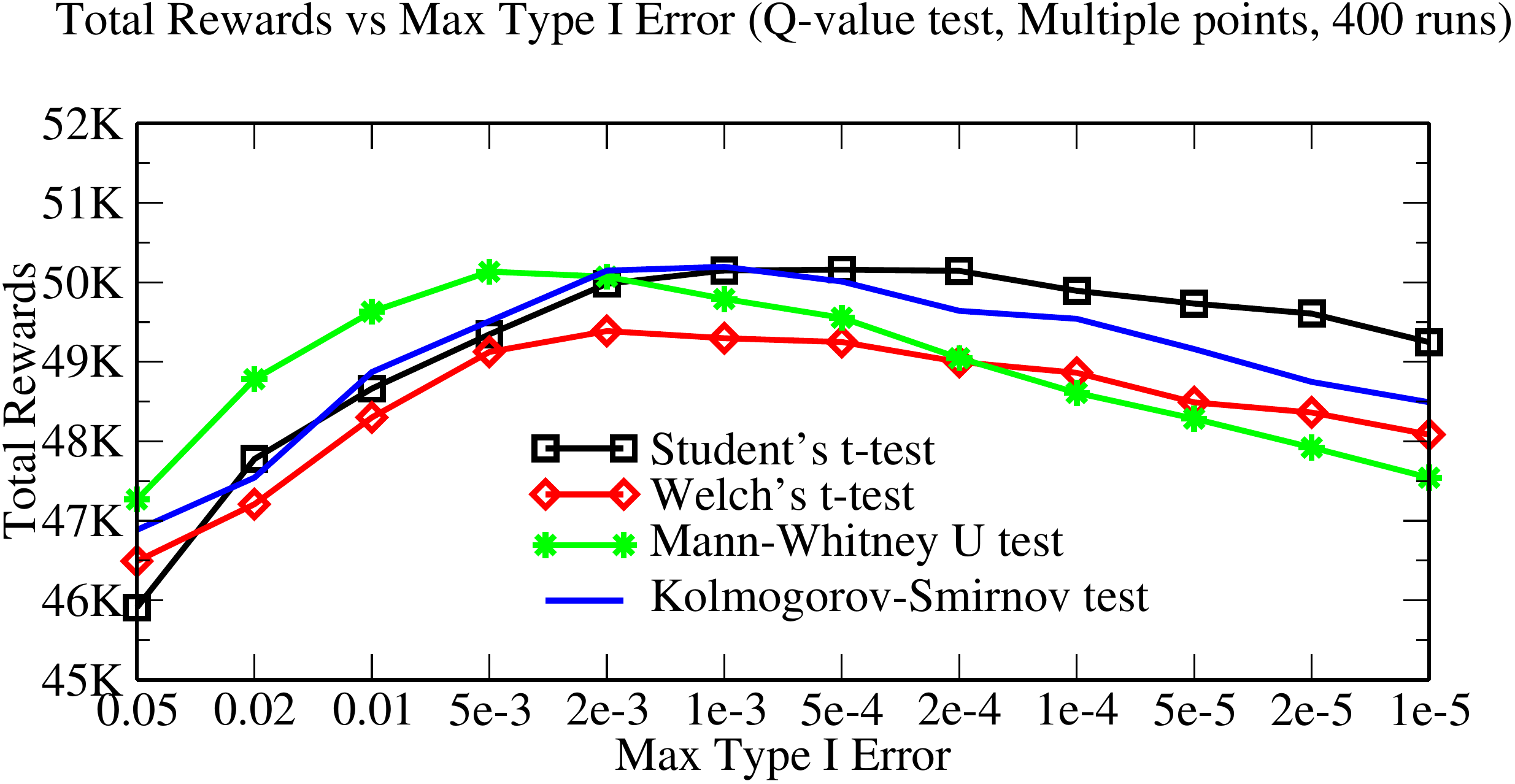}
		\vspace{-0.2in} 
        \caption{Q-value test, Multiple splitting points}
    \end{subfigure}
    \caption{Performance of the four statistical tests for three different splitting criteria as a
             function of the maximum error}
    \label{fig:stat-performance}
		\vspace{-0.15in} 
\end{figure*}

For all statistical tests and splitting criteria, 
the typically used margin of 0.05 (corresponding to $95\%$ confidence) results in a very large number
of incorrect decisions (Figure \ref{fig:stat-accuracy}). An incorrect decision is performed when the test decides to split a non-relevant random dimension apart from the cluster size, incoming load and read percentage. 
In order to effectively restrict those incorrect decisions, the margin 
needs to be set significantly lower, in the area of $0.002$ (or even lower for the Q-value test
on multiple points). If set low enough though, in most cases the mistakes were completely avoided,
especially when using splitting criteria that only consider a single splitting point per
parameter (Parameter test and Q-value test at the median point). The only exception to
this is Welch's test in the case of the Q-value test with multiple splitting points, where even
with a very strict margin of $0.00001$ it only managed to achieve an approximately 95\% accuracy. 

All the criteria achieve their lowest accuracy when using the \emph{Q-value test with multiple splitting
points}. The reason that the other two criteria have higher accuracy, is that they consider only a single splitting
point per parameter, which divides almost equally the available experiences, ending up
comparing sets of approximately equal sizes. However, the criterion \emph{Q-value test with multiple splitting points} considers splitting points that leave only a handful of experiences in each set, making the decision significantly
more difficult. The statistical test that was affected the most by this fact was \emph{Welch's test}, since it
assumes that the two populations have different standard deviations, and as a result
is very easily misled if one of the two groups has very few elements.

Comparing the accuracy of the statistical tests between the two criteria that consider a single splitting point
per parameter, the two tests that do not assume a normal distribution of the values (\emph{Mann-Whitney U test} and \emph{Kolmogorov-Smirnov test}) both achieve a better accuracy if they run on the values of the parameters.
However, the two tests that do assume a normal distribution (\emph{Student's t-test} and \emph{Welch's test}) had their accuracy very noticeably reduced if they run on the values of the parameters. This is explained
by the fact that the values of some of the parameters are discrete. In such cases these tests give misleading estimation of the standard deviation of the populations. 

Overall, the \emph{Mann-Whitney U test} achieves the highest accuracy for all the criteria. The
\emph{Kolmogorov-Smirnov test} is second in all the criteria, as long as the error margin is strict
enough, while on the contrary, it performs very poorly if the margin is lenient $max\_type\_I\_error$ values of 0.002 or higher).
Finally, the two tests assuming normal distributions, \emph{Student's t-test} and \emph{Welch's test},
perform well only if they run on the Q-values (which are not discrete), and consider a single
splitting point. 

The disadvantage of selecting a very small error margin is that it greatly reduces
the total amount of splits performed by all tests (Figure \ref{fig:stat-splits}).
For the criterion \emph{Q-value test on the median point} the amount of splits for all tests is very close.
However, for the \emph{Parameter test} the two tests that assume normal
distributions (which are also the least accurate on this criterion) performed significantly
more splits. Finally, when using the \emph{Q-value test with multiple splitting points}, as expected, all the tests increase the number of splits performed. 
Among the statistical tests, the \emph{Mann-Whitney U test}, which is the most accurate, performs the
smallest amount of splits in all situations, and is the least affected by the consideration of multiple
splitting points.

In terms of performance of the algorithm (i.e. the success of the algorithm in taking optimal decisions), the typical value of $0.05$ for the achieved significantly suboptimal performance with all criteria and statistical tests (Figure \ref{fig:stat-performance}). 
The \emph{Mann-Whitney U test}, being the most accurate one, manages to perform very well with all the
criteria. However, performing the smallest amount of splits, it achieves that performance for
bigger values of the error margin, in the area of $0.005$. The \emph{Kolmogorov-Smirnov
test} also does well with all the criteria, achieving its best performance for a slightly
smaller value of $0.002$.

\begin{figure*}[t!]
    \centering
        \includegraphics[width=0.95\textwidth]{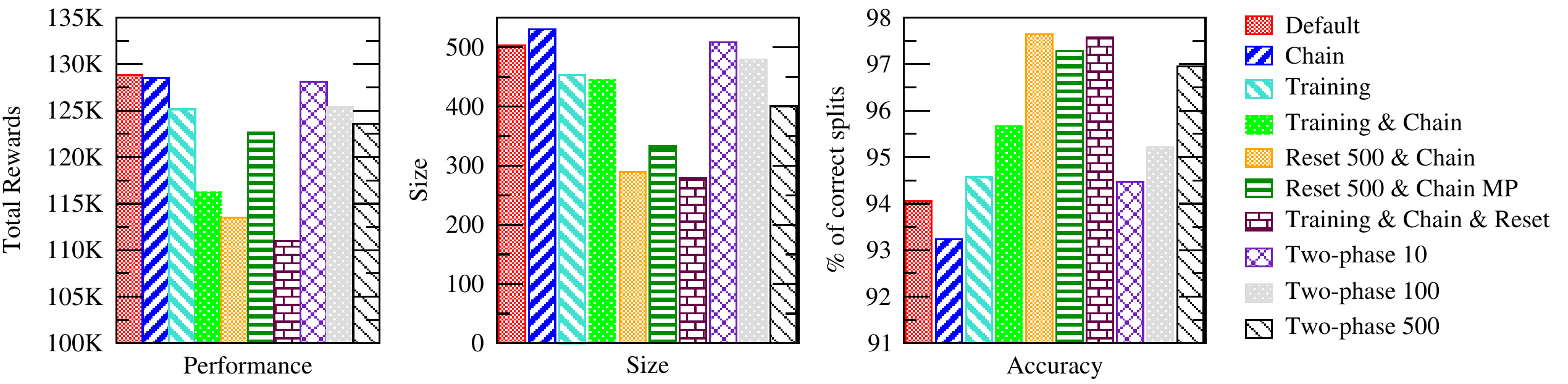}
        \vspace{-0.15in}
   \caption{Performance, decision tree size and accuracy for different splitting strategies (200 runs)}
    \label{fig:splitting-strategy}\vspace{-0.15in}
\end{figure*}

As mentioned, the difference between the equal variance t-test 
(\emph{Student's t-test}) and its unequal variance counterpart (\emph{Welch's t-test}) is that the
\emph{Student's t-test} makes the assumption that the variances of the two populations are
equal, while \emph{Welch's t-test} is expected to be more accurate if this does not 
necessarily hold \cite{welch1, welch2}.
In cases that the two compared variances are unequal, the \emph{Welch's t-test} is more applicable.
For example, this holds if certain values of a parameter cause instability of the system; then the variance of the Q-values in that part of the state space would be expected to have a higher variance. Oppositely, the assumption
of equal variances increases the strength of the \emph{Student's t-test}; in our experiments
we expect the variances of the population to be approximately equal more often than not. This is the reason that the \emph{Student's t-test} clearly outperforms the \emph{Welch's t-test}, often by a significant margin.
Compared to the other tests, the \emph{Student's t-test} does very well with the \emph{Q-value test} criterion, if
the tested values are continuous. However, the performance of the \emph{Student's t-test} with the \emph{Parameter test} criterion drops significantly below that of the \emph{Mann-Whitney test} and \emph{Kolmogorov-Smirnov test}.

\subsection{Splitting Criteria}\label{sec:simulation_splitcriteria}

\begin{figure}[t!]
\includegraphics[width=0.5\textwidth]{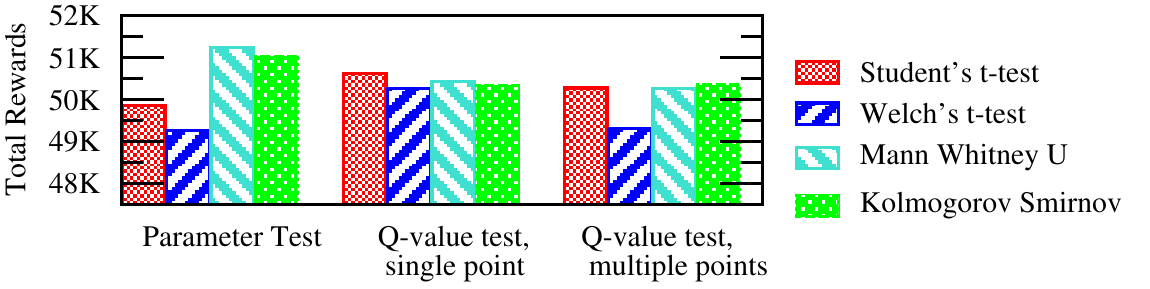}
 \vspace{-0.1in}
    \caption{Performance comparison of all the splitting criteria using their optimal settings (400 runs)}\vspace{-0.15in}
    \label{fig:criteria}
   
\end{figure}

In this experiment we compare the performance of the splitting criteria under the
optimal settings derived from the previous experiments (Figure \ref{fig:criteria}).
The two tests that do not assume a normal distribution of the values, the \emph{Kolmogorov-Smirnov test}
and the \emph{Mann Whitney U test}, perform better employed in the \emph{Parameter test} criterion, while the
two tests that assume normal distributions perform better employed in the \emph{Q-value test} criterion. 
The reason is that some of the parameters are discrete; this means that their distribution differs significantly from a normal
distribution, resulting in lower performance of the tests if they are applied on the values of these parameters.
Oppositely, these two tests perform better employed in the \emph{Q-value test} criterion, since the Q-values are generally not discrete.

Among all options, the \emph{Mann Whitney U test} achieved the best performance, employed in the \emph{Parameter test}. We note that this test is the most
accurate in terms of the number of incorrect splits, as shown in Section
\ref{sec:statistical-significance}.
Finally, between the two available options for the \emph{Q-value test}, namely considering a single or multiple
splitting points, the consideration of a single splitting point achieves generally better results, while it produces smaller decision trees.

\subsection{Splitting Strategy}\label{sec:simulation_splitstrategy}

\eat{
By default, we attempt to perform a split on the starting state of each experience after the
experience has been acquired. In this experiment we test the performance of different approaches 
on this decision. 
One such approach, whose aim is to accelerate the growth of the tree, is to attempt to
split every state in the model, regardless of whether it was involved in an experience. This makes
sense because the splitting criteria take into account the values of other states, and so it is 
possible that a change in the value of one state triggers a split in another. If a split is indeed
performed on any state, we repeat the process. We call this procedure a \textit{Chain Split}.

One other strategy is to delay the splitting until a significant amount of data have been acquired. 
The reasoning behind this is that once more data are available the splitting criteria may be able
to make better decisions and build a better decision tree compared to the one built with less data.
Additionally, we experimented with periodically resetting the decision tree and performing a 
\textit{Chain Split} to rebuild it.

Finally, we tested the splitting strategy used in \cite{continuousutree}. This included
splitting the algorithm in two phases, a \textit{Data Gathering} phase where data are
collected but no splits are performed, and a \textit{Processing Phase}, where all the nodes
of the model are tested one by one to check if a split is needed, and if so, perform the
splits. We tested performing this Processing Phase every 10, 100 and 500 steps.
}

In this experiment we compare various different splitting strategies. Mixing the three basic strategies defined in Section \ref{sec:splitstrategy}, we create the following variety of strategies:
\begin{enumerate}{\leftmargin=1em}
        \setlength{\topsep}{0pt}
        \setlength{\parskip}{0pt}
        \setlength{\partopsep}{0pt}
        \setlength{\parsep}{0pt}         
        \setlength{\itemsep}{0pt} 
\item \textit{Default:} Attempt to split the starting state for each new experience.
\item \textit{Chain:} Perform a \textit{Chain Split} with every new experience.
\item \textit{Training:} Allow splitting to begin after a training of 2500 steps and then start splitting with every new experience.
\item \textit{Training \& Chain:} Allow splitting to begin after a training of 5000 steps,
and also perform one chain split at that time; then continue splitting with every new experience.
\item \textit{Reset 500 \& Chain:} At each step perform a \emph{Chain Split} and every 500 steps reset the decision tree. 
\item \textit{Reset 500 \& Chain MP:} As above, but using the multiple points
\emph{Q-value test} criterion, attempting to split each state at multiple points per parameter.
\item \textit{Training \& Chain \& Reset:} After a training of 5000 steps, perform a \emph{Chain Split} and then do the same by resetting the tree every 500 steps.
\item \textit{Two-phase 10:} After a \emph{Data Gathering} phase of 10 steps, run a \emph{Processing} phase.
\item \textit{Two-phase 100:} After a \emph{Data Gathering} phase of 100 steps, run a \emph{Processing} phase.
\item \textit{Two-phase 500:} After a \emph{Data Gathering} phase of 500 steps, run a \emph{Processing} phase.
\end{enumerate}

In Figure \ref{fig:splitting-strategy} we present the performance in terms of total rewards, the size of the produced decision tree and the accuracy for every different strategy. Even though \emph{Chain Split} adopts a much more aggressive (and computationally intensive)  strategy in attempting to grow the decision tree, it has a performance similar to the \emph{Default} strategy. Also, even though \emph{Chain Split} performs 30 additional splits on average, the split quality decreases. The relatively low amount of additional splits reveals that the default strategy already depletes most of the opportunities to create new states.

Waiting for more data to be available in order to start splitting did not perform well. Despite 
offering a slight increase in the accuracy of the splits, (in the order of 1-2\%), it causes a
10\% reduction in their number and in the case of strategy (iv) a drop in performance. 

Periodically resetting the tree to rebuild it provides the most
accurate splits on the final tree. This is expected since the splits are performed with the
maximum amount of data. However, the resulting tree size is significantly smaller in this
case, limiting the performance of this strategy.


The results for strategy (vii) show the impact of having a long training before starting splitting, then doing a chain split, and resetting the decision tree every 500 steps thereafter. During the long training the optimal action in each region of the state space is repeated only a few times due to the $e$-greedy strategy. Therefore, less data is available in order to perform splits, which results in a smaller tree, with consequences in performance.

Finally, using a \emph{Data Gathering} and a \emph{Processing} phase periodically instead of regularly splitting
performed better the smaller that period was. If \emph{Processing} is performed every 10 steps it nearly reached
the performance of the default strategy (though having a significantly larger running time), but for periods larger than that it falls behind. 

Overall, the results of this experiment show that the default method of attempting to
split the initial state of each experience is both efficient and effective. 

\subsection{Initial Size of the Decision Tree}

\begin{figure}[t!]
\includegraphics[width=0.5\textwidth]{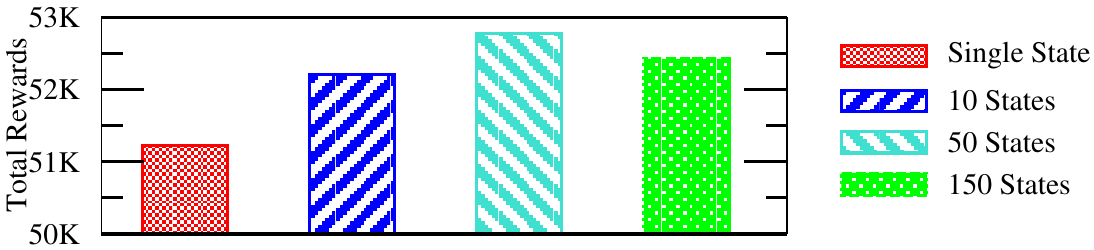} \vspace{-0.1in}
    \caption{The effect of starting with an existing decision tree on performance}\vspace{-0.15in}
    \label{parameters1}
\end{figure}

\begin{figure}[t!]
\includegraphics[width=0.5\textwidth]{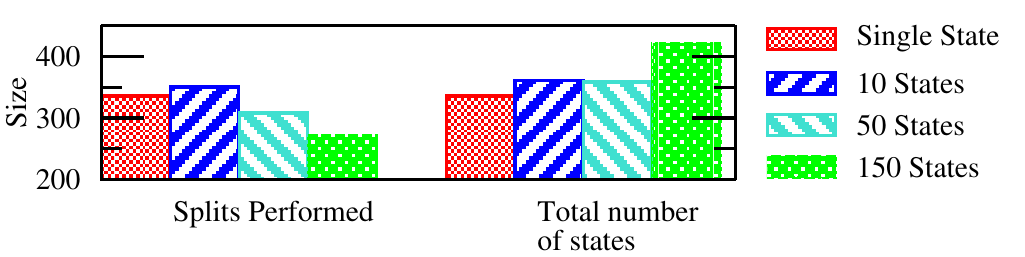} \vspace{-0.1in}
    \caption{The number of splits and the final number of states as a function of the initial size of the decision tree}\vspace{-0.15in}
    \label{parameters2}
\end{figure}


Even though the decision tree algorithms are designed to work on large and unknown state spaces, in
some applications there is partial knowledge about the state space beforehand. For example, in our problem setting, we expect that the size of the cluster is always an important parameter in the elasticity decisions taken by the RL agent.
Therefore, as an extension to the splitting mechanism we also enable splitting on multiple points, instead of only one. Multiple splitting points allows for the easy and efficient construction of pre-defined decision trees. 
If a node is split in multiple points, the additional states are appended at the end of the $state$ vector, and changes in the $transitions$, $rewards$ and $experiences$ vectors are performed accordingly.

We experiment with the size of the decision tree at the beginning of the execution of \emph{MDP\_DT}. Instead of starting with a single state, we create trees that are equivalent
to 1-dimensional, 2-dimensional and 3-dimensional grids.
%
As expected, this pre-partitioning of the state space improves the performance of the algorithm
(Figure \ref{parameters1}). Moreover, in the case of the 1-dimensional grid, the algorithm even performs more splits. (Figure \ref{parameters2}): since the initial tree only contains 10 states (leaving still a lot of room for additional splits), and
the resulting subspaces are easier to be handled by the algorithm. On the contrary, adding a third dimension decreases the performance of the algorithm, even though there is still room to perform a large number of splits. This is a clear indication that the state space in this problem can be partitioned more efficiently than an orthogonal grid.

Overall, our results show that the ability to begin with a predefined partitioning of the state space and dynamically increase the resolution during the algorithm execution can increase the performance of the algorithm.


\section{Experimental Results} \label{sec:experimental}

In this section we evaluate our theoretical and simulation-based findings on a real experimental setup. We employ our RL techniques in order to dynamically scale a distributed database cluster deployed in a cloud environment under varying workloads. We evaluate the performance of \emph{MDP\_DT} using different training data sizes and workloads, we compare \emph{MDP\_DT} with model free and static partitioning algorithms and finally we showcase \emph{MDP\_DT'}s ability to benefit from multiple parameters and perform correct decisions.

\subsection{System and Algorithm Setup} \label{sec:exp-setup}

In order to test our proposal in a real cloud environment, we use an \textit{HBase 1.1.2} NoSQL
distributed database cluster running on top of \textit{Hadoop 2.5.2}.
We generate a mix of different read and write intensive workloads of varying amplitude by utilizing the \textit{YCSB} \cite{cooper2010benchmarking} benchmark, while \textit{Ganglia} \cite{massie2004ganglia}
is used for the collection of the NoSQL cluster metrics. The cluster runs on a private \textit{OpenStack}
IaaS cloud setup. The coordination of the cluster is performed by a modified version of \cite{tiramola1}.
In any case, we allow for 5 different actions, which include adding or removing 1 or 2 VMs from
the cluster, or doing nothing. Decisions are taken every 15 minutes, as it is found that this time is necessary for the system to reach a steady state after every reconfiguration. This fact also confirms our full-model choice, since this time is adequate to calculate the updated model after every decision. The cluster size used in our experiments ranges between 4 and 15 VMs. Each VM in
the HBase cluster has 1GB of RAM, 10GB of storage space and 1 virtual CPU, while the
master node has 4GB of RAM, 10GB of storage and 4 virtual CPU's.
For the training of the decision tree based models we use a set of 12 parameters including:
\begin{myitemize}
\item The cluster size
\item The amount of RAM per VM
\item The percentage of free RAM
\item The number of virtual CPU's per VM
\item The CPU utilization
\item The storage capacity per VM
\item The number of I/O requests per second
\item The CPU time spent waiting for I/O operations
\item A linear prediction of the next incoming load (equal to two times the current load minus
the last recorded load)
\item The percentage of read requests in the queries
\item The average latency of the queries
\item The network utilization
\end{myitemize}
For the Decision Tree based algorithms we split the state space using the Mann Whitney U test with the Parameter Test splitting criterion over a default splitting strategy, following our findings in sections \ref{sec:simulation_splitcriteria} and \ref{sec:simulation_splitstrategy}. The model-based algorithms update their optimal policies utilizing the \emph{Prioritized Sweeping Algorithm} \cite{prioritized-sweeping}. For the static partitioning algorithms we select two dimensions that were found to be the most relevant for the cluster performance (i.e., the cluster size and the linear load prediction) divided in 12 and 8 equal partitions respectively, resulting in 96 states. We note that this setup is optimal for the static schemes, as they require a small number of relevant states to behave correctly.

\subsection{MDP\_DT Behavior}

\begin{figure*}[ht!]
    \centering
    \begin{subfigure}[b]{0.32\textwidth}
        \includegraphics[width=\textwidth]{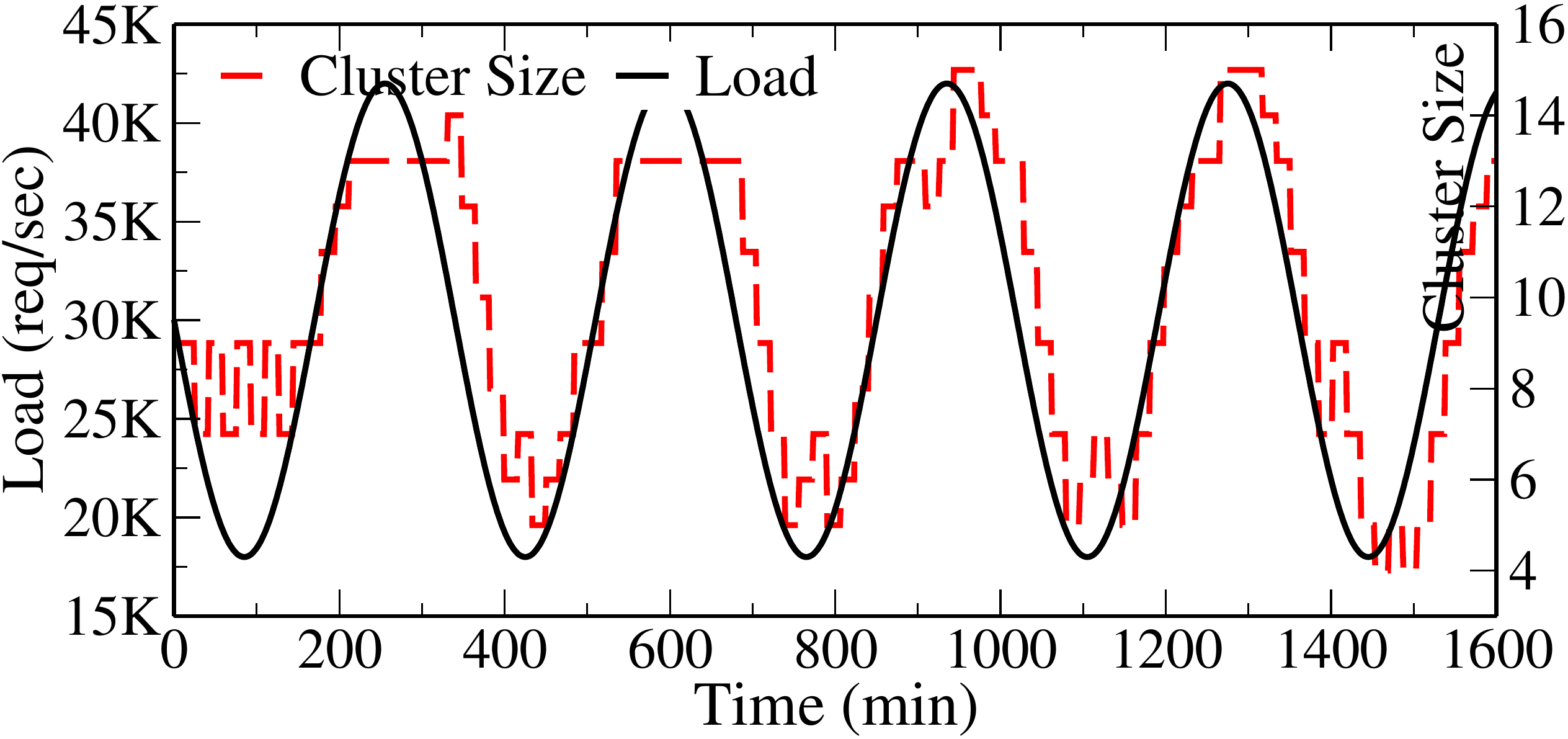}
		\vspace{-0.2in}        
        \caption{Sinusoidal load (minimal dataset)}
        \label{img:tiramola-1k}
    \end{subfigure}
    ~
    \begin{subfigure}[b]{0.32\textwidth}
        \includegraphics[width=\textwidth]{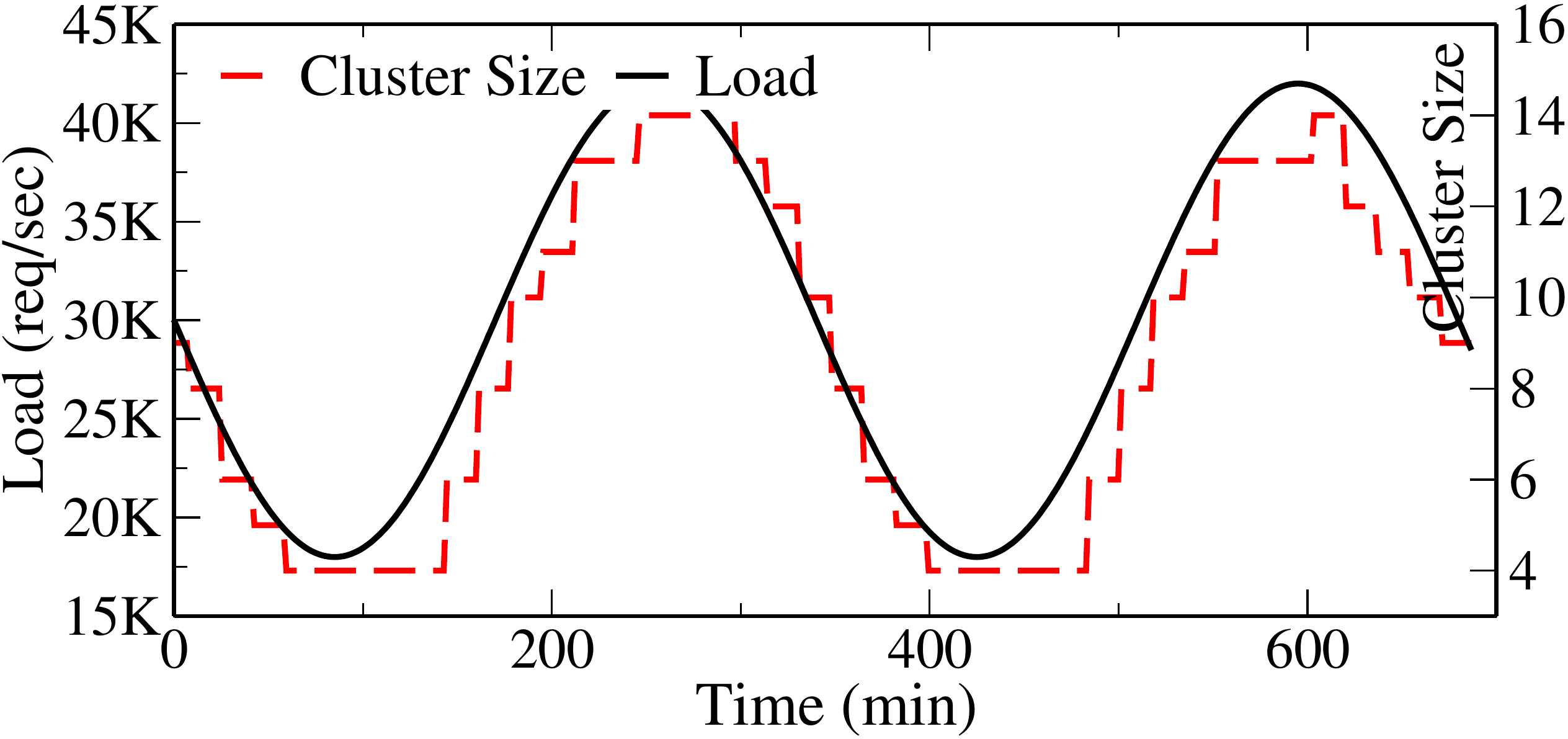}
		\vspace{-0.2in}        
        \caption{Sinusoidal load (small dataset)}
        \label{img:tiramola-5k}
    \end{subfigure}
    ~
    \begin{subfigure}[b]{0.32\textwidth}
        \includegraphics[width=\textwidth]{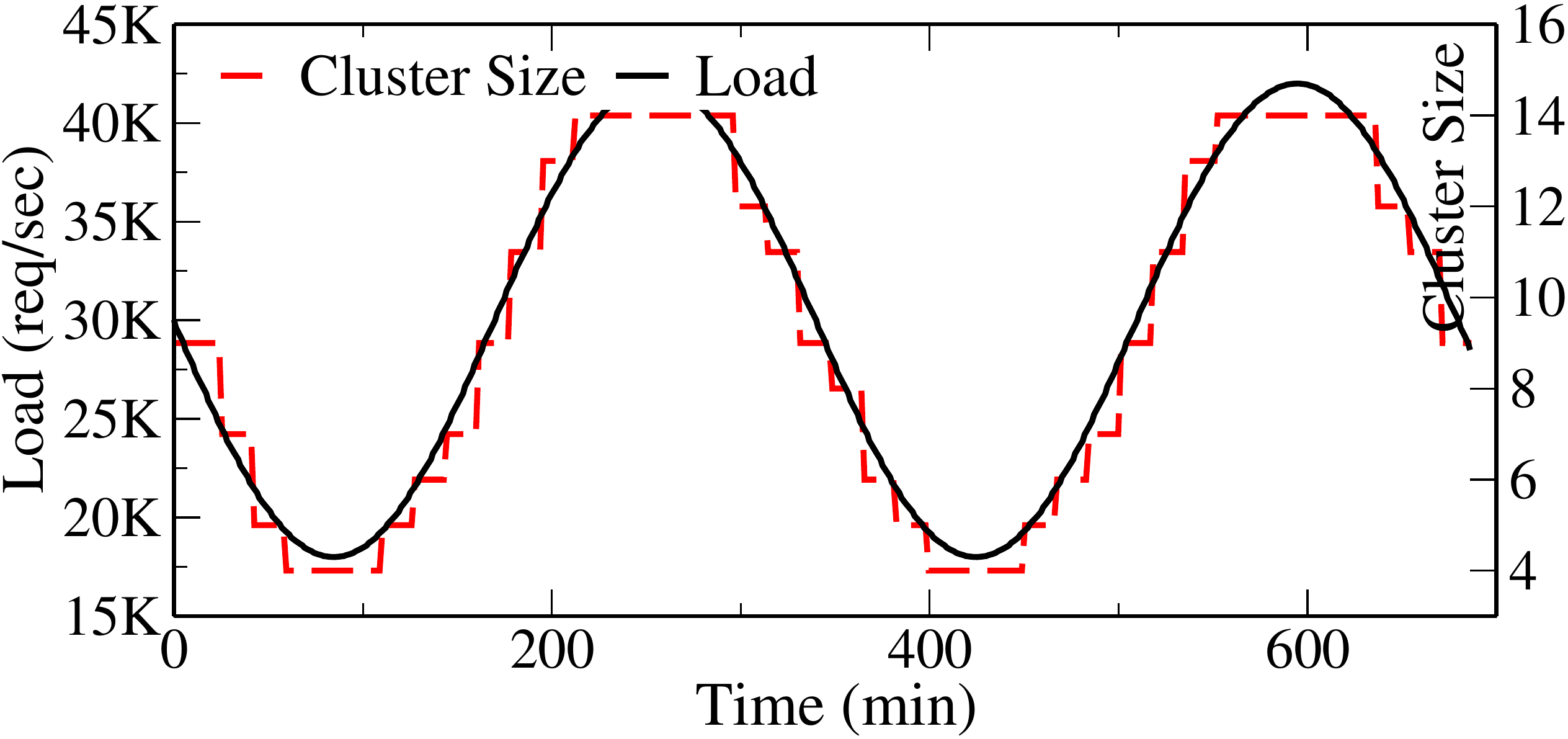}
		\vspace{-0.2in}        
        \caption{Sinusoidal load (large dataset)}
        \label{img:tiramola-20k}
    \end{subfigure}
    ~
    \begin{subfigure}[b]{0.32\textwidth}
        \includegraphics[width=\textwidth]{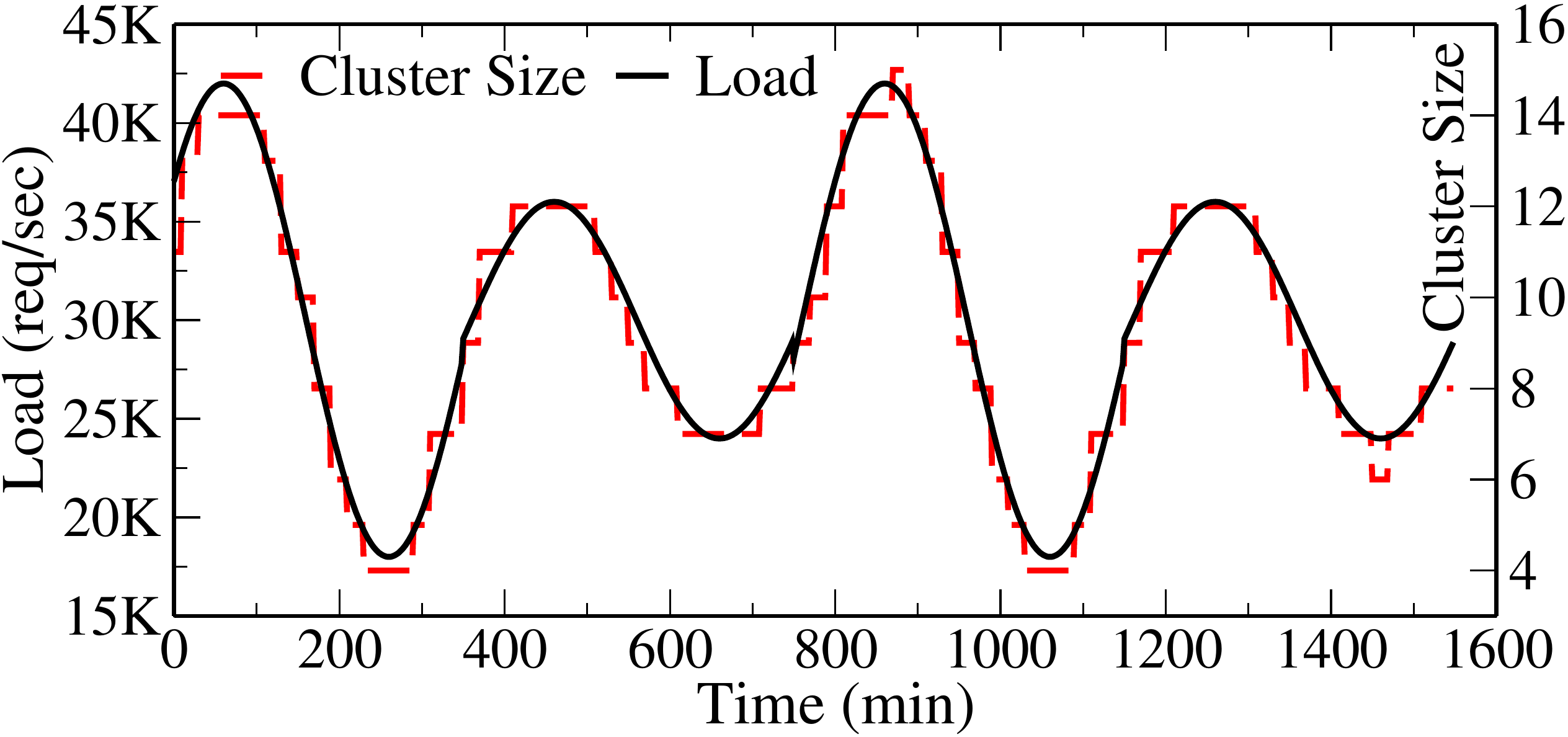}
		\vspace{-0.2in}        
        \caption{Sinusoidal load with alt. amplitude}
        \label{img:alt-ampl}
    \end{subfigure}
    ~
    \begin{subfigure}[b]{0.32\textwidth}
        \includegraphics[width=\textwidth]{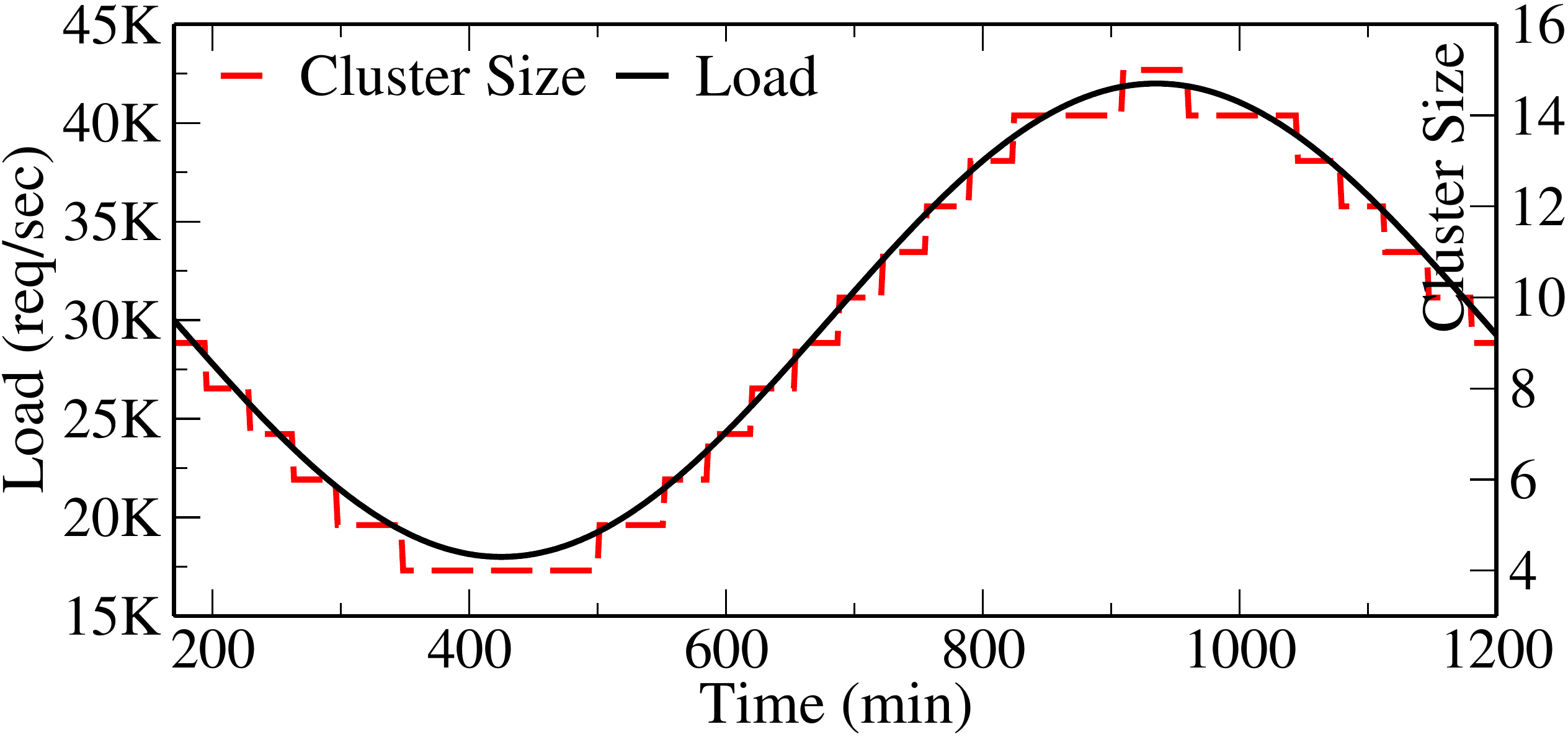}
		\vspace{-0.2in}        
        \caption{Slow sinusoidal load}
        \label{img:slow-sin}
    \end{subfigure}
    ~
    \begin{subfigure}[b]{0.32\textwidth}
        \includegraphics[width=\textwidth]{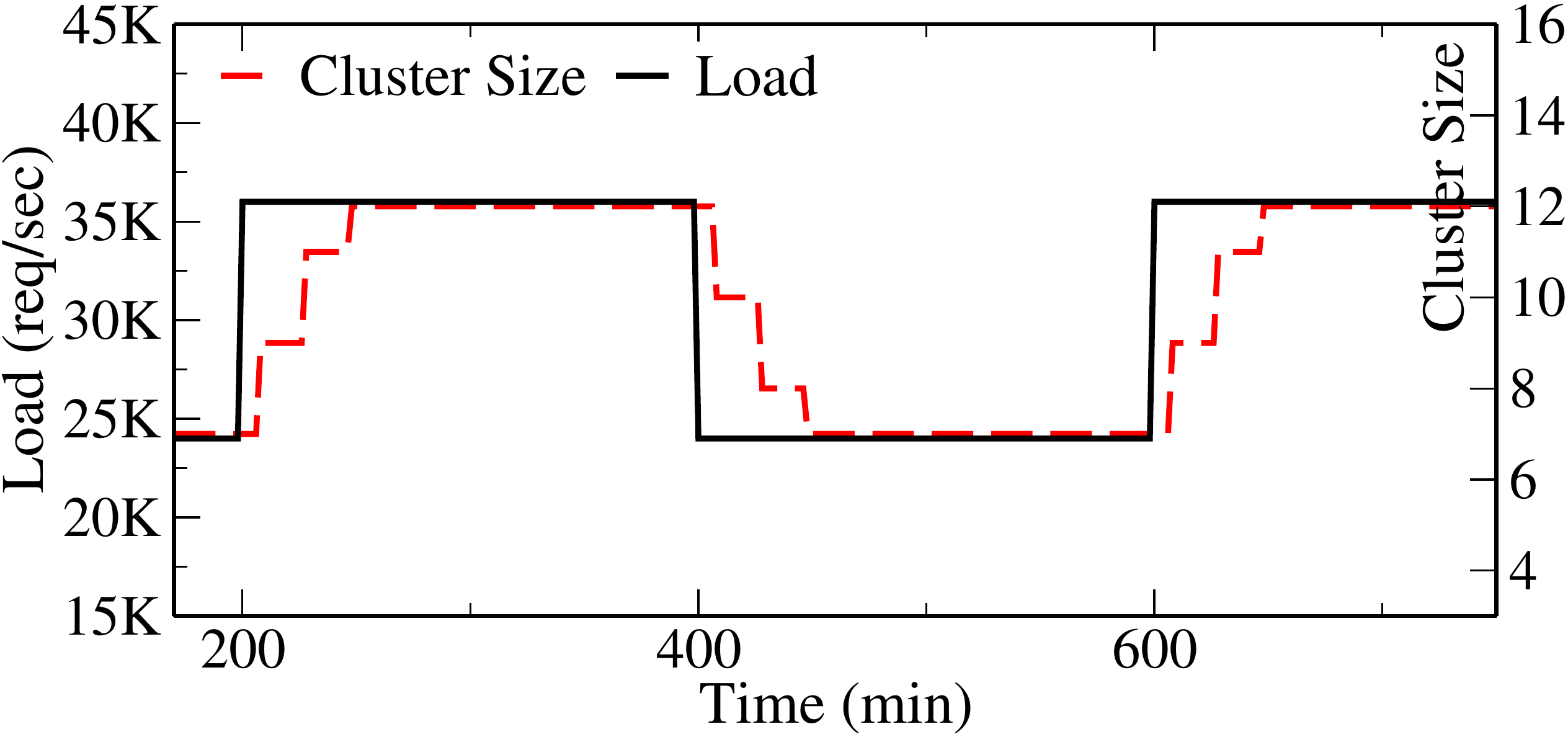}
		\vspace{-0.2in}        
        \caption{Square pulse load.}
        \label{img:pulse-load}
    \end{subfigure}
		\vspace{-0.1in}        
    \caption{\emph{MDP\_DT} under different training data (Figs \ref{img:tiramola-1k}, \ref{img:tiramola-5k} and \ref{img:tiramola-20k}) and workload types (Figs \ref{img:alt-ampl}, \ref{img:slow-sin} and \ref{img:pulse-load}).}
    \label{img:mdp_dteval}
    \vspace{-0.1in}
\end{figure*}

\begin{figure*}[ht!]
    \centering
    \begin{subfigure}[b]{0.23\textwidth}
        \includegraphics[width=\textwidth]{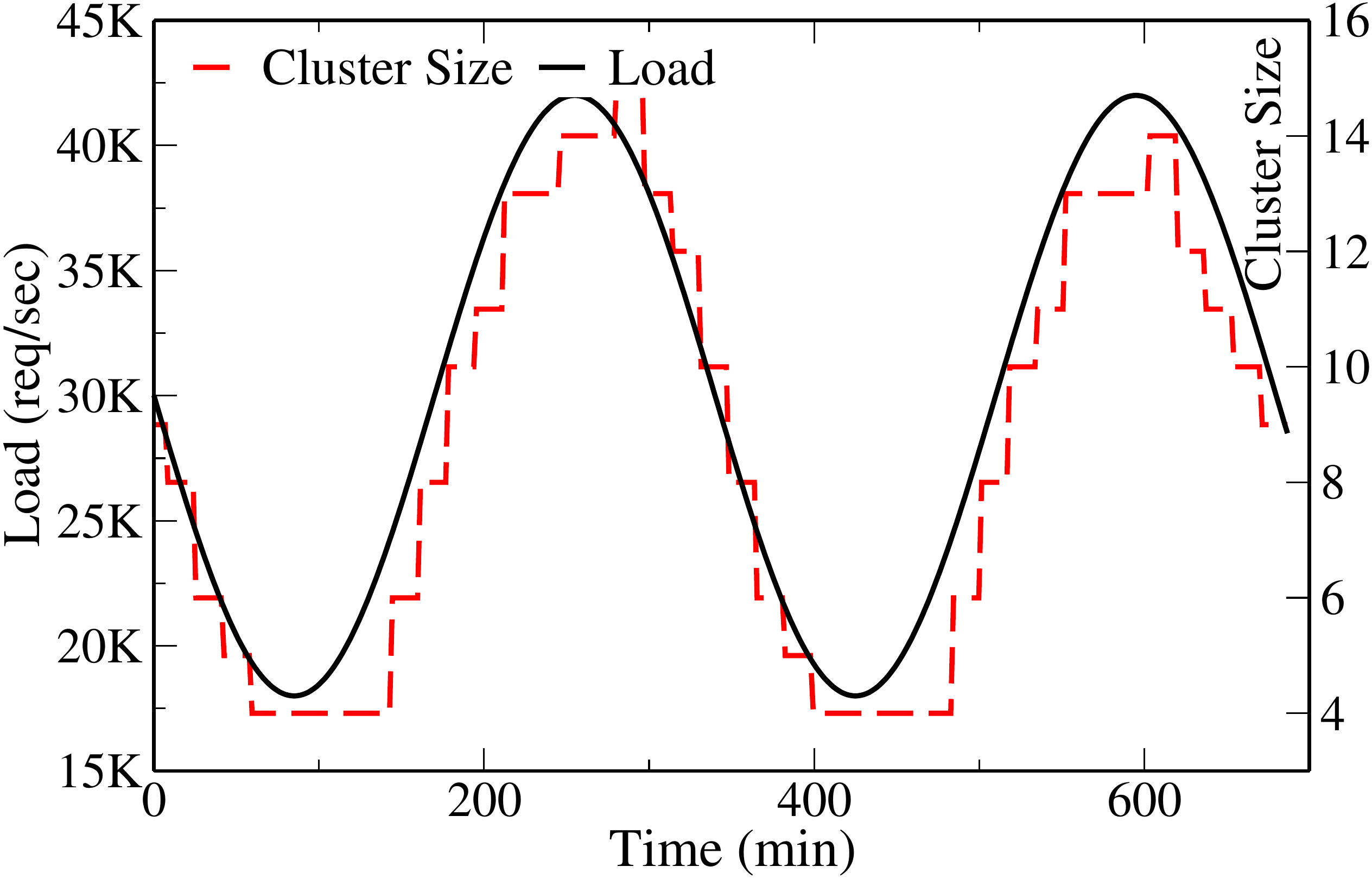}
		\vspace{-0.2in}        
        \caption{MDP\_DT, small dataset}
        \label{img:mdpdtsmall}
    \end{subfigure}
    ~
    \begin{subfigure}[b]{0.23\textwidth}
        \includegraphics[width=\textwidth]{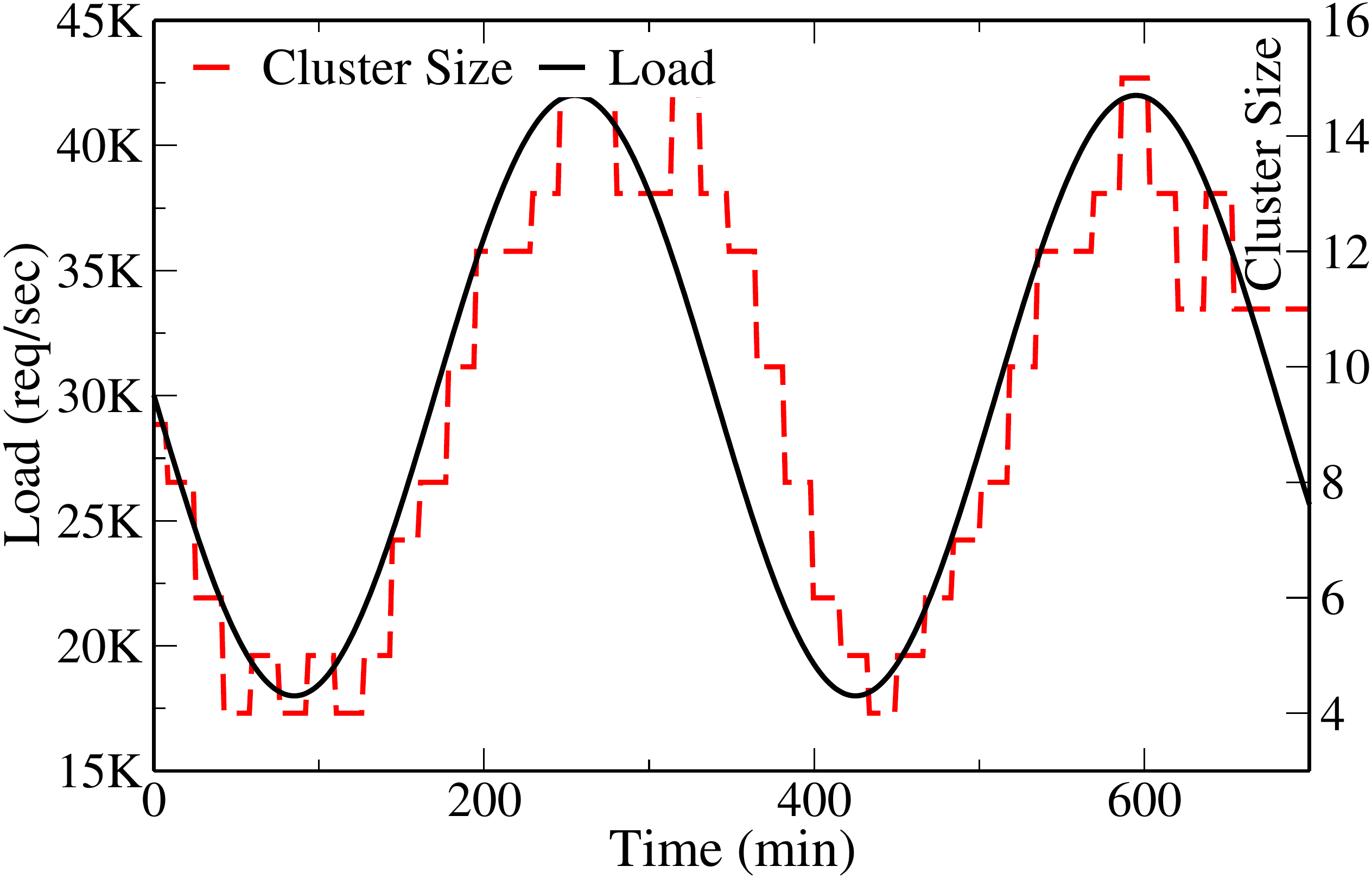}
		\vspace{-0.2in}        
        \caption{MDP, small dataset}
        \label{img:mdpsmall}
    \end{subfigure}
    ~
    \begin{subfigure}[b]{0.23\textwidth}
        \includegraphics[width=\textwidth]{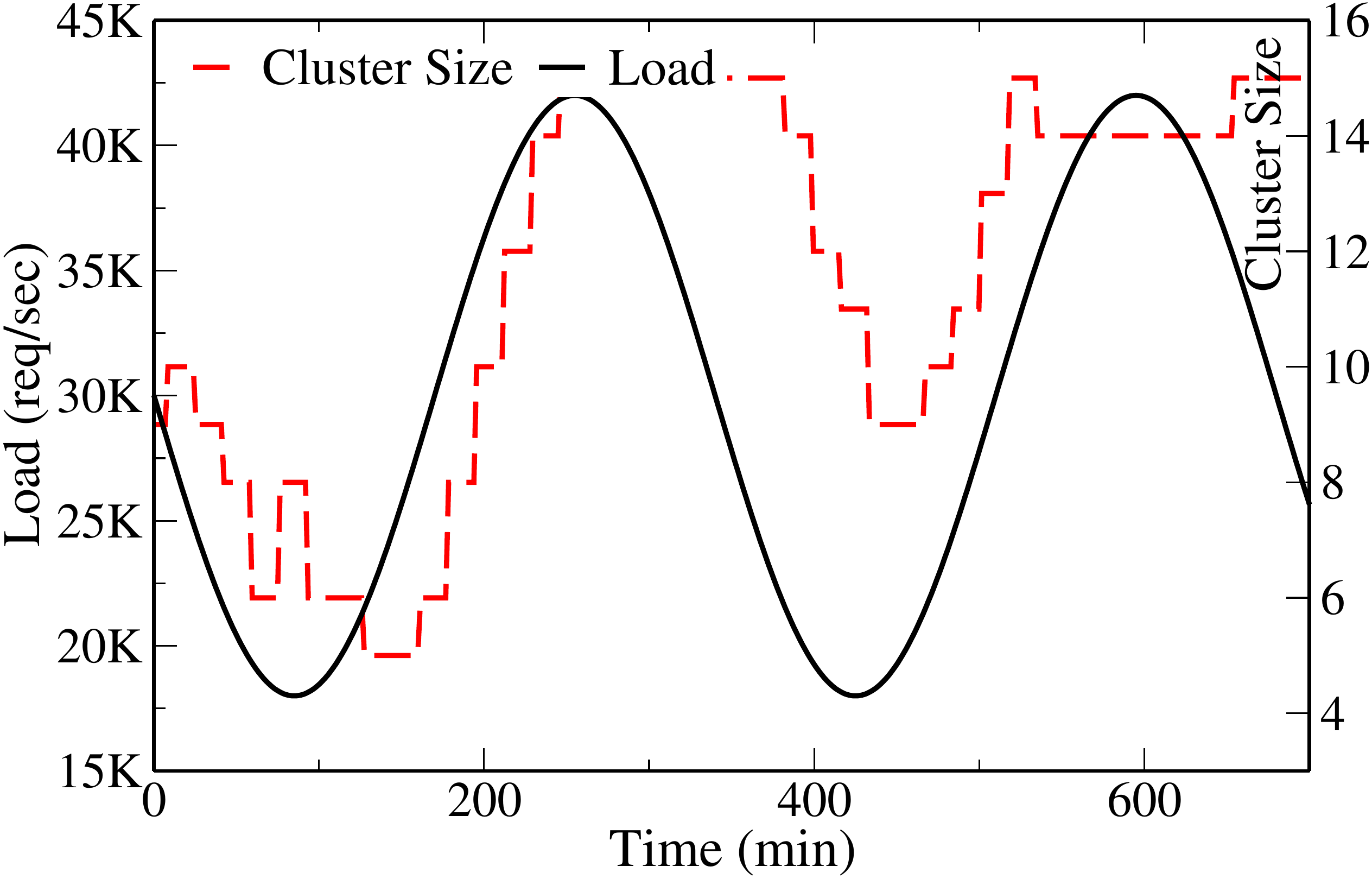}
		\vspace{-0.2in}        
        \caption{QDT, small dataset}
        \label{img:qdtsmall}
    \end{subfigure}
    ~
    \begin{subfigure}[b]{0.23\textwidth}
        \includegraphics[width=\textwidth]{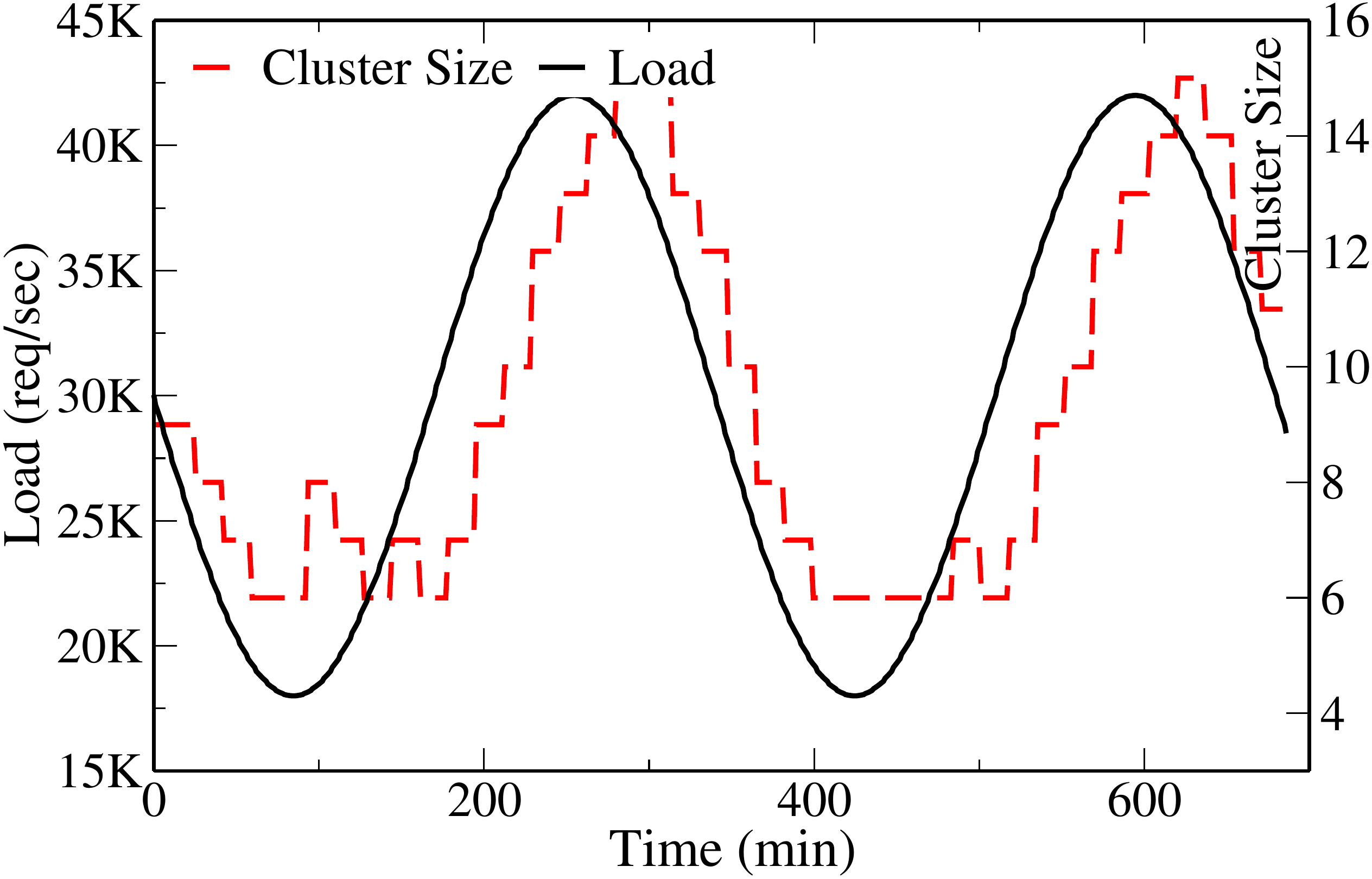}
		\vspace{-0.2in}        
        \caption{Q-learning, small dataset}
        \label{img:qsmall}
    \end{subfigure}
    ~
    \begin{subfigure}[b]{0.23\textwidth}
        \includegraphics[width=\textwidth]{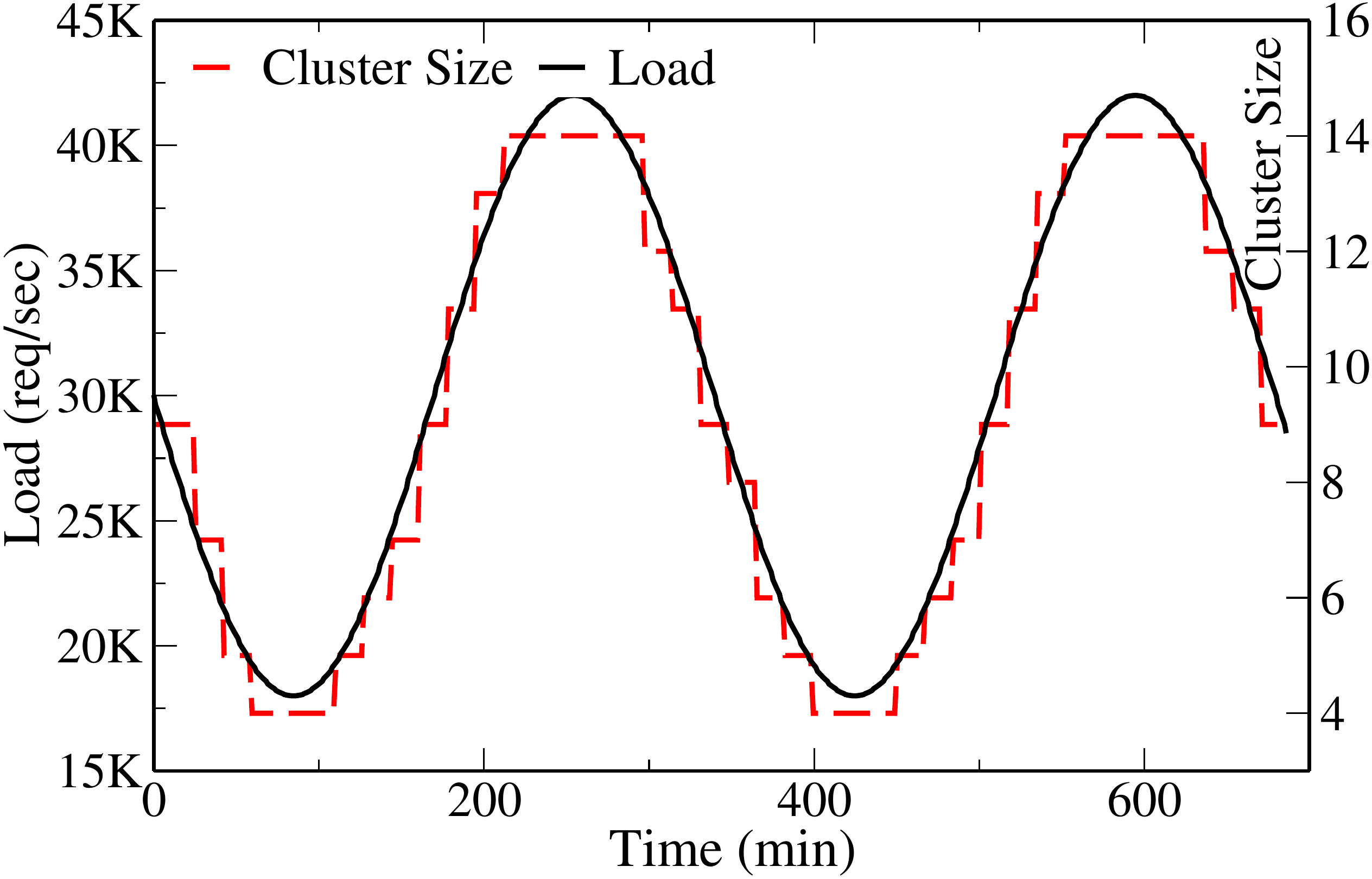}
		\vspace{-0.2in}        
        \caption{MDP\_DT, large dataset}
       \label{img:mdpdtlarge}
    \end{subfigure}
    ~
    \begin{subfigure}[b]{0.23\textwidth}
        \includegraphics[width=\textwidth]{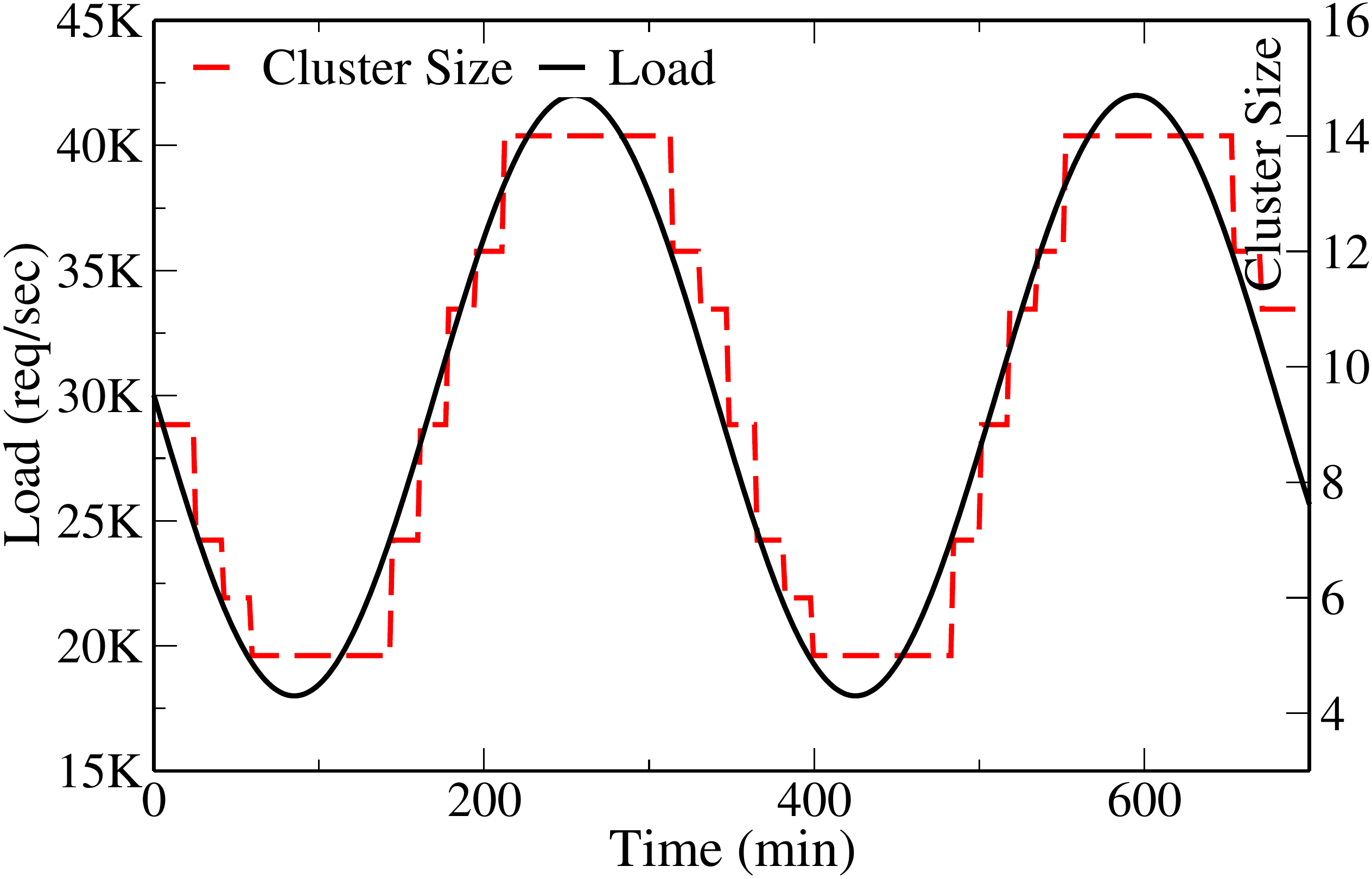}
		\vspace{-0.2in}        
        \caption{MDP, large dataset}
        \label{img:mdplarge}
    \end{subfigure}
    ~
    \begin{subfigure}[b]{0.23\textwidth}
        \includegraphics[width=\textwidth]{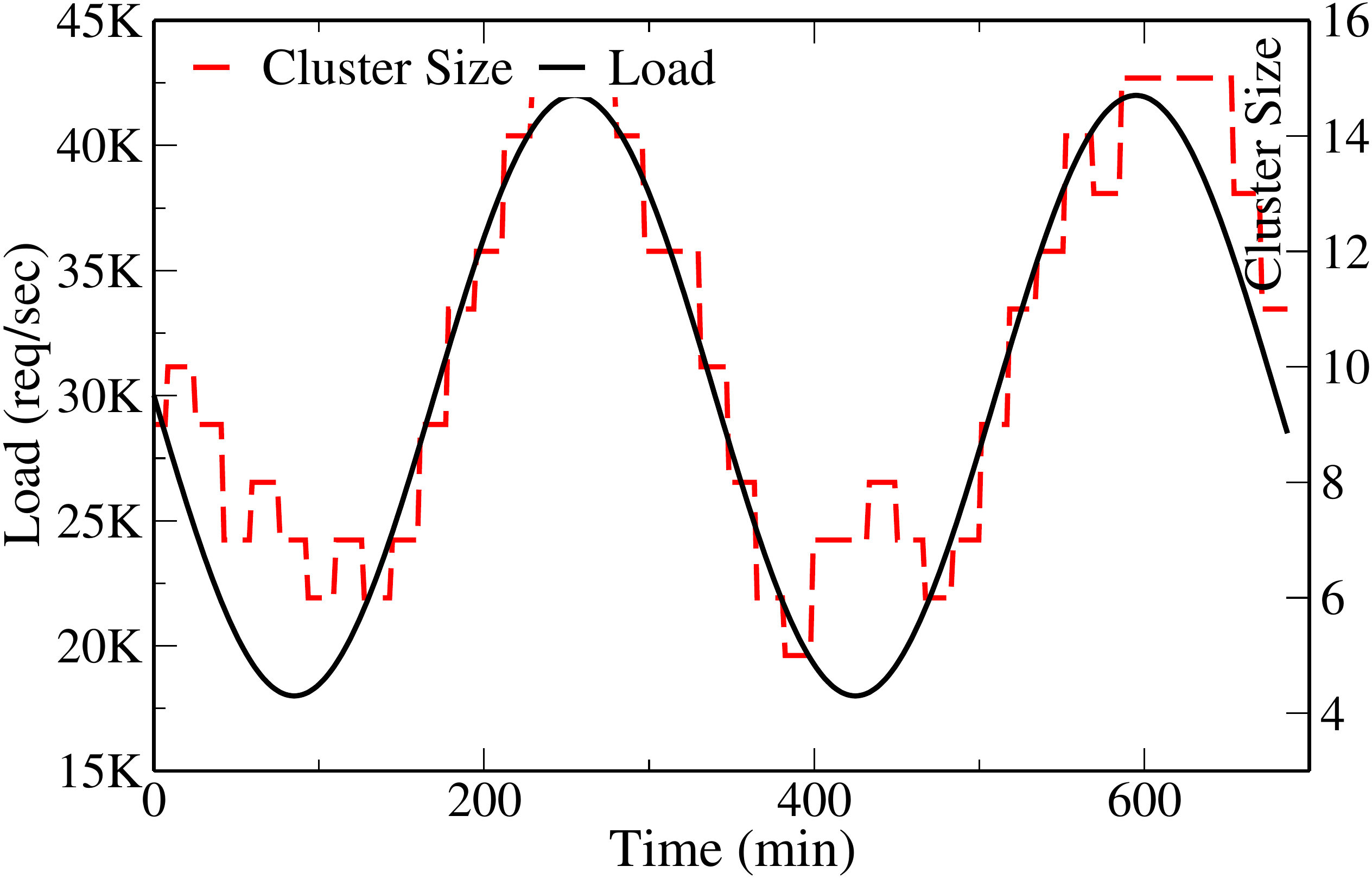}
		\vspace{-0.2in}        
        \caption{QDT, large dataset}
        \label{img:qdtlarge}
    \end{subfigure}
    ~
    \begin{subfigure}[b]{0.23\textwidth}
        \includegraphics[width=\textwidth]{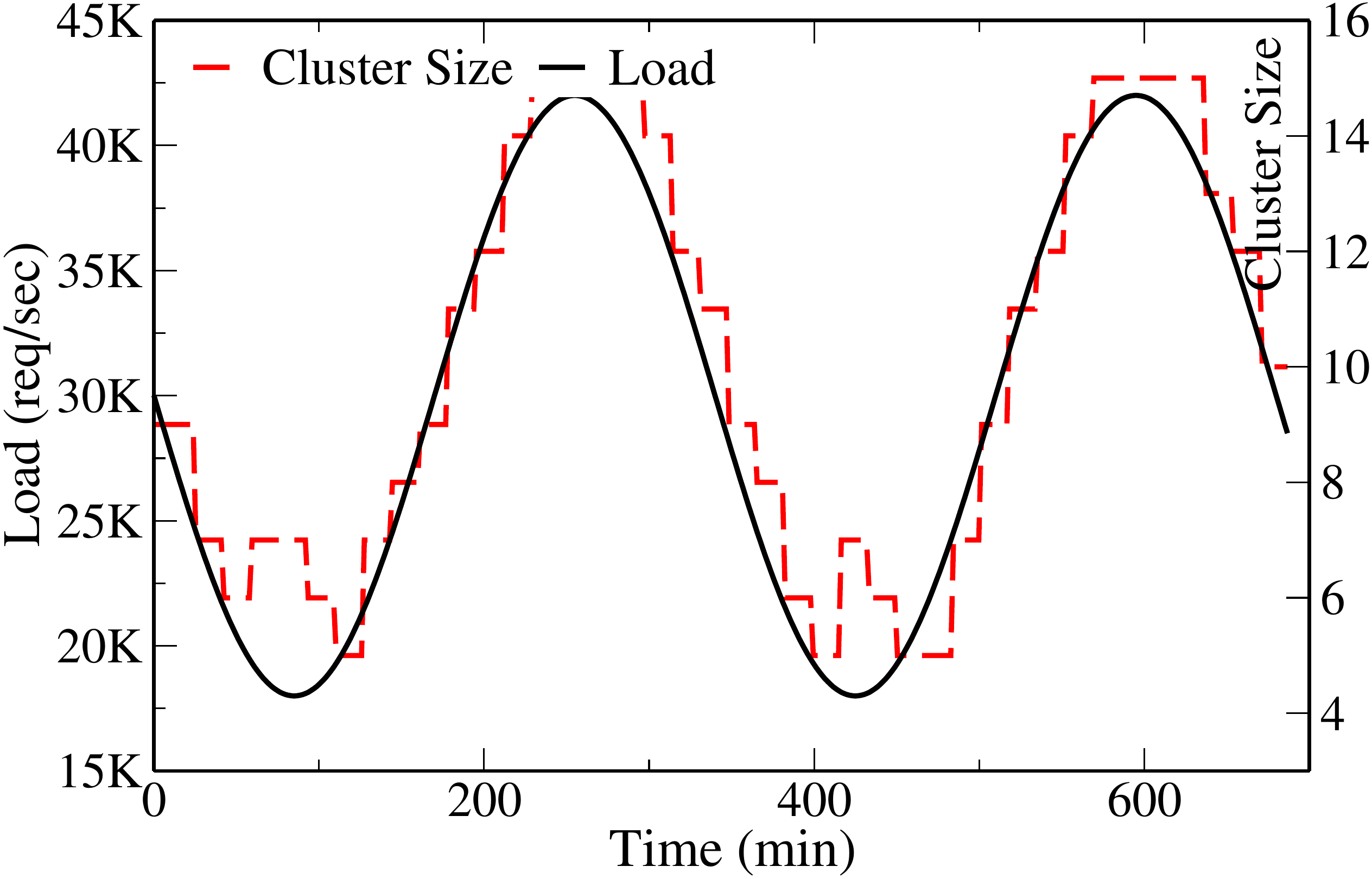}
		\vspace{-0.2in}        
        \caption{Q-learning, large dataset}
        \label{img:qlarge}        
    \end{subfigure}
    \caption{Comparison of the behavior of model-based vs model-free and decision-tree based vs static algorithms.}
    \label{img:exp-models}
\end{figure*}

In this section we test \emph{MDP\_DT} using different workloads and training set sizes. In Figure \ref{img:mdp_dteval} we present our findings. Every experiment runs for a total of 700-1200 minutes (X axis). The solid line represents the workload in terms of Reqs/sec (left Y axis) whereas the dotted line represents the cluster size (right Y axis). Every step in the dotted line represents an \emph{MDP\_DT} action of adding or removing VMs.

We initialize the \emph{MDP\_DT} decision tree with 6 states and let it partition the state space
on its own from that point on. The training load is a sinusoidal load of varying
amplitude. 

First, we run \emph{MDP\_DT} with a minimal dataset of 500 experiences (Figure \ref{img:tiramola-1k}).
When trained with this dataset, only 17 splits are performed during the training (4 using
the size of the cluster and 13 using the incoming load), increasing the total number of states
to 22. During this run 12 additional splits are performed (4 using cluster size, 
7 using incoming load and 1 using latency),
allowing \emph{MDP\_DT} to continuously adapt and follow the incoming load (Figure \ref{img:tiramola-1k}).
When provided with bigger datasets of 1000 and 20000 experiences, the performance improves and
very closely converges to the incoming load, ending up with 54 and 576 states respectively
(Figures \ref{img:tiramola-5k} and \ref{img:tiramola-20k}). Finally, in Figures \ref{img:alt-ampl}, \ref{img:slow-sin} and \ref{img:pulse-load} we can observe how \emph{MDP\_DT} adapts to arbitrary workloads that has not previously encountered.

\subsection{Using Different Algorithms}

\begin{figure*}[t!]
    \centering
    \begin{subfigure}[b]{0.32\textwidth}
        \includegraphics[width=\textwidth]{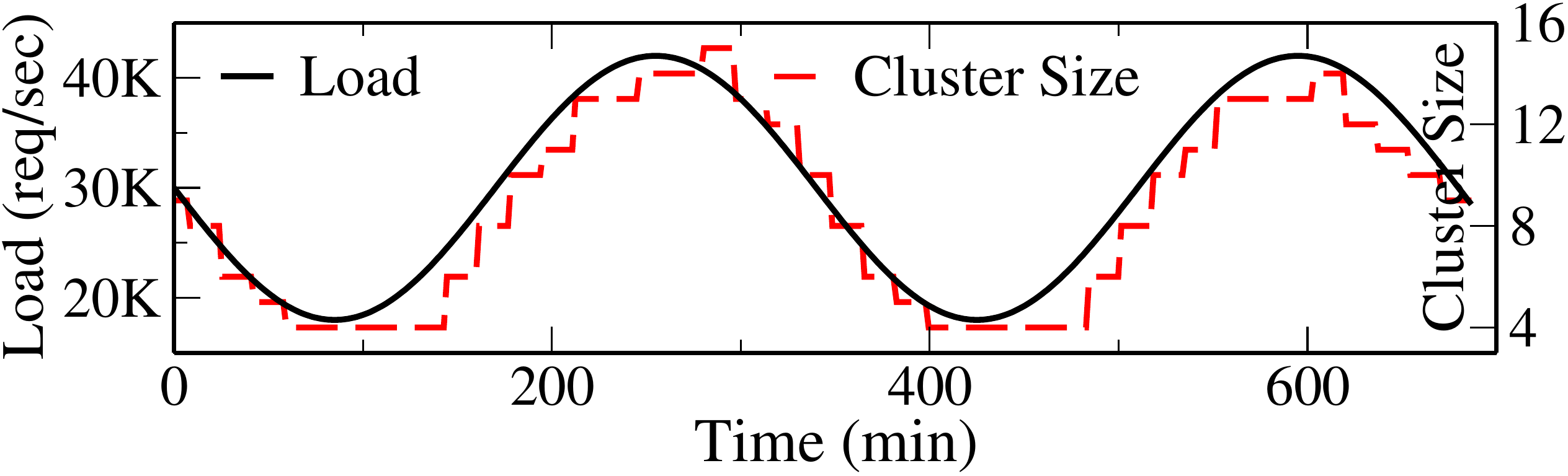}
		\vspace{-0.2in}        
        \caption{All parameters}
        \label{img:allparams}
    \end{subfigure}
    ~
    \begin{subfigure}[b]{0.32\textwidth}
        \includegraphics[width=\textwidth]{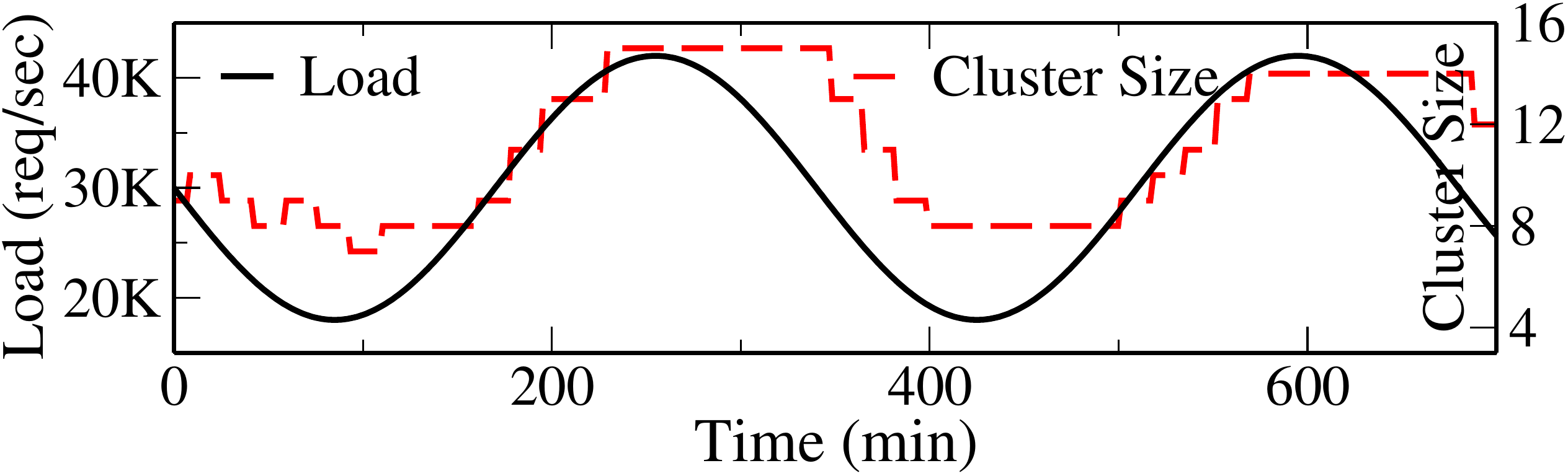}
		\vspace{-0.2in}        
        \caption{Average latency}
        \label{img:onlylat}
    \end{subfigure}
	~
    \begin{subfigure}[b]{0.32\textwidth}
        \includegraphics[width=\textwidth]{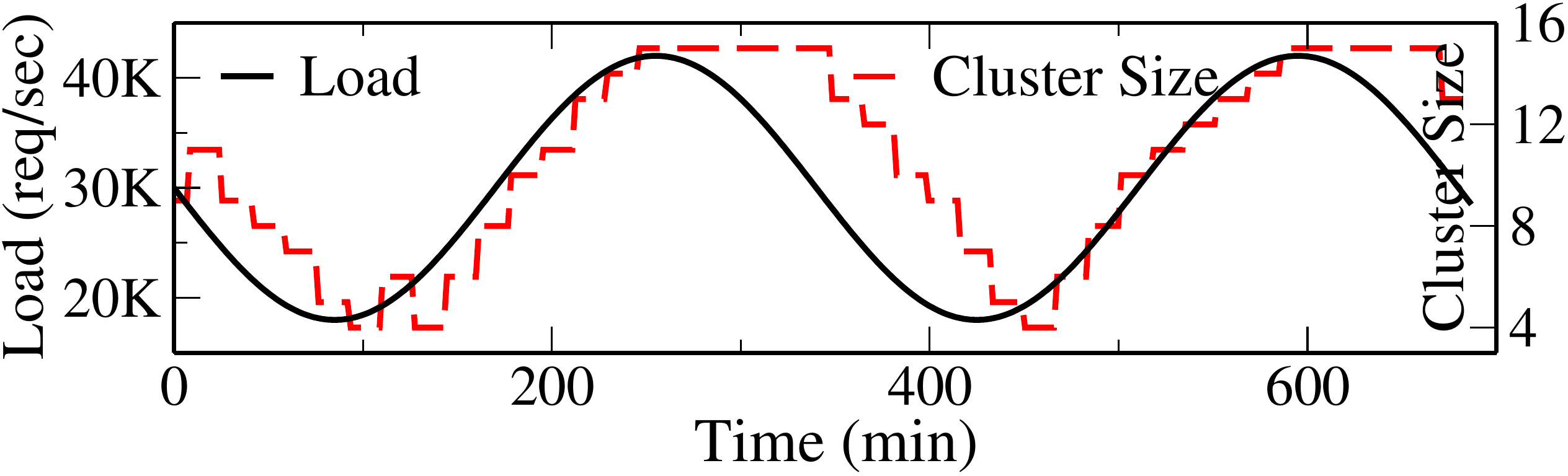}
		\vspace{-0.2in}        
        \caption{CPU utilization}
        \label{img:onlycpu}
    \end{subfigure}
    ~
    \begin{subfigure}[b]{0.32\textwidth}
        \includegraphics[width=\textwidth]{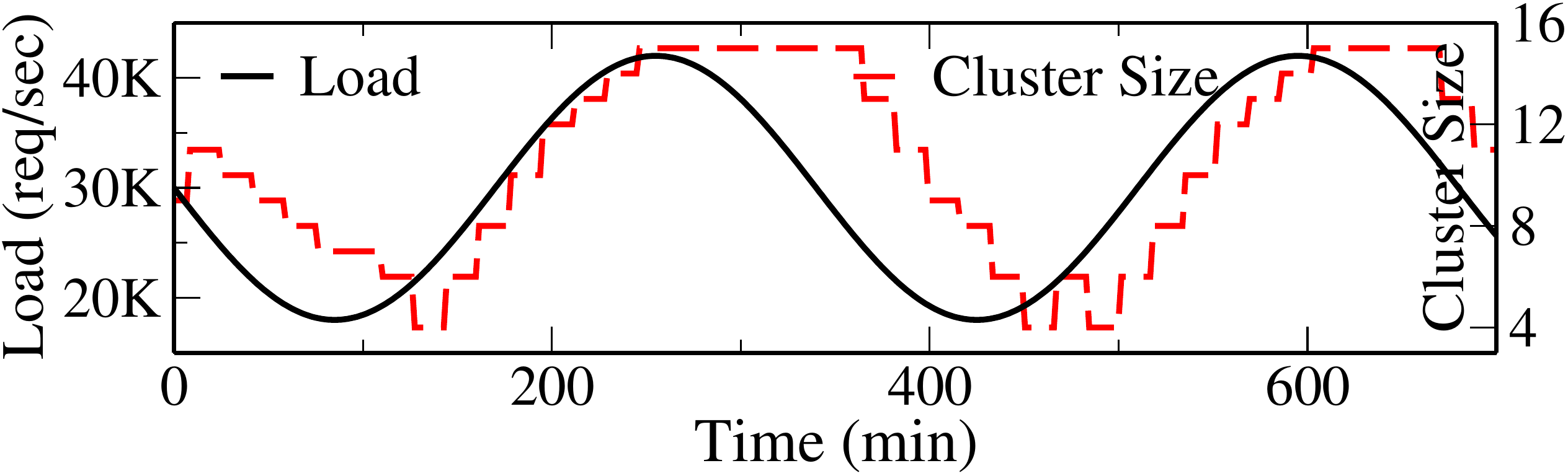}
		\vspace{-0.2in}        
        \caption{One min load}
        \label{img:onlyload}
    \end{subfigure}
	~
    \begin{subfigure}[b]{0.32\textwidth}
        \includegraphics[width=\textwidth]{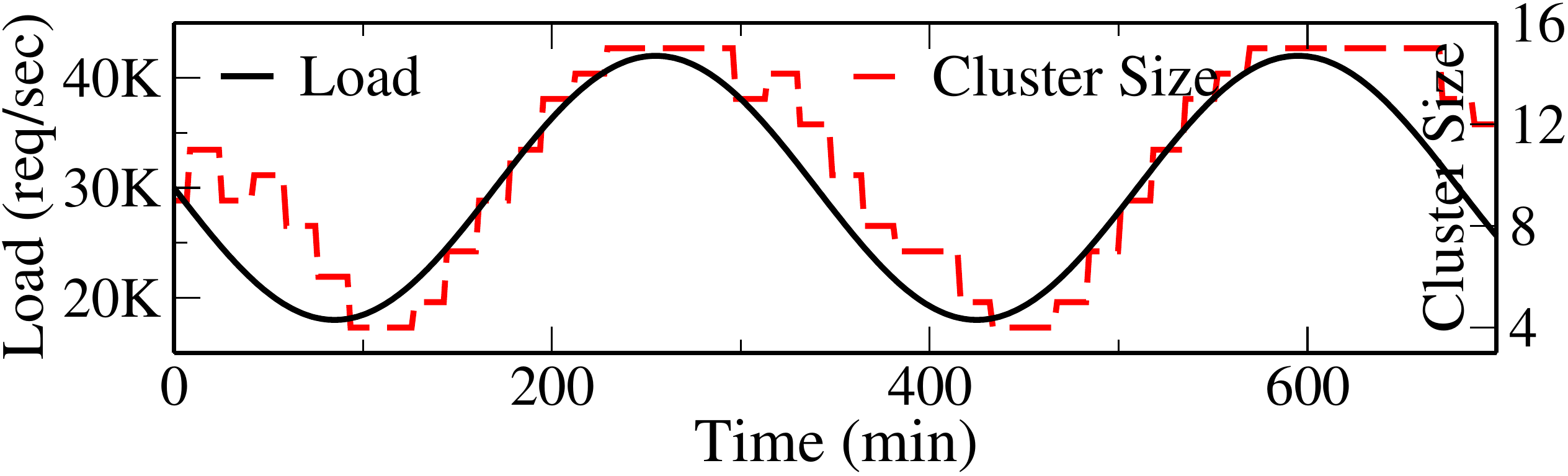}
		\vspace{-0.2in}        
        \caption{Network Usage}
        \label{img:onlynet}
    \end{subfigure}
    ~
    \begin{subfigure}[b]{0.32\textwidth}
        \includegraphics[width=\textwidth]{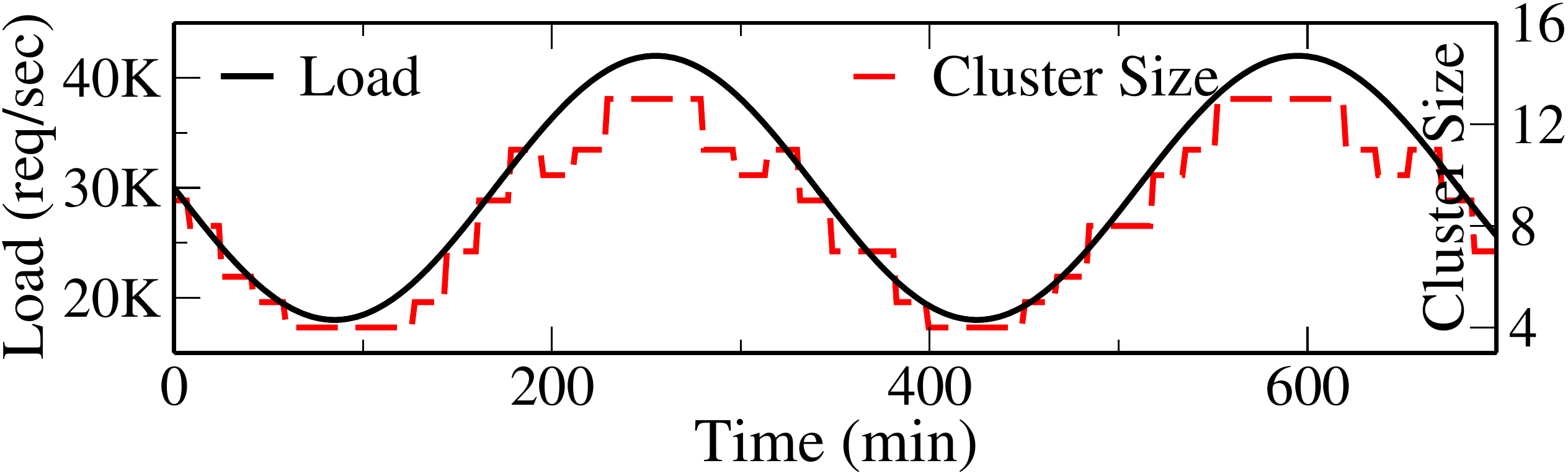}
		\vspace{-0.2in}        
        \caption{Load prediction}
        \label{img:onlyloadpred}        
    \end{subfigure}
		\vspace{-0.15in}        
    \caption{System behavior when allowing splits with only the cluster size plus 
             one additional parameter}
    \label{fig:exp-parameters}
\end{figure*}

In this experiment we test the system's behavior compared to model-free (i.e., \emph{Q-learning}) and static partitioning schemes. 
The combinations of these schemes lead to four different algorithms, namely the model-based adaptively partitioned \emph{MDP\_DT}, the model-based statically partitioned \emph{MDP}, and the respective model-free versions (\emph{Q\_DT} and \emph{Q-learning}). 
We present our findings in Figure \ref{img:exp-models}.
We notice that \emph{MDP\_DT} follows the applied workload very closely (Figures \ref{img:mdpdtsmall} and \ref{img:mdpdtlarge}), with a better result when a large training set is utilized (Figure \ref{img:mdpdtlarge}). Indeed the algorithm's decisions over time depicted in the red dotted line seem to perfectly adapt to the observed workload, since both lines almost overlap.
The full-model based \emph{MDP} algorithm also performs well in this setting (Figure \ref{img:mdplarge}), managing to
follow the incoming load reasonably well even when trained with the small dataset (Figure \ref{img:mdpsmall}).
At the same time, it manages to perform accurately when trained with more data.
This is not a surprise since this problem has a reasonably simple state space,
and a partitioning using only the size of the cluster and the incoming load is quite sufficient
to capture the behavior in this experiment: in this ideal setting the state space is very accurately defined and also a lot of training data is available. Yet, in sudden load spikes observed in the max and min load values it takes more time to respond compared to the \emph{MDP\_DT} case. 

The \emph{Q-learning} based algorithms though both require a large amount of
data to follow the incoming load effectively (Figures \ref{img:qdtsmall}, \ref{img:qsmall}, \ref{img:qdtlarge} and \ref{img:qlarge}). 
In this experiment the decision tree based \emph{Q-learning} algorithm (\emph{QDT})
achieves the weakest performance with the small dataset (Figure \ref{img:qdtsmall}).
With this few data this is not totally unexpected, since
at the start of the training that model uses the first data it acquires
to perform splits, but then discards it after the splits have
been performed, leaving it with very little available information to make decisions.
If more training data is provided though, it catches up to the 
traditional \emph{Q-learning} model (Figure \ref{img:qdtlarge}).
However, they both are noticeably less stable (they cannot follow the observed load in an adequate manner) compared to the full model approaches.

\subsection{Restricting the Splitting Parameters}


\begin{figure}[t!]
\centering\includegraphics[width=0.5\textwidth]{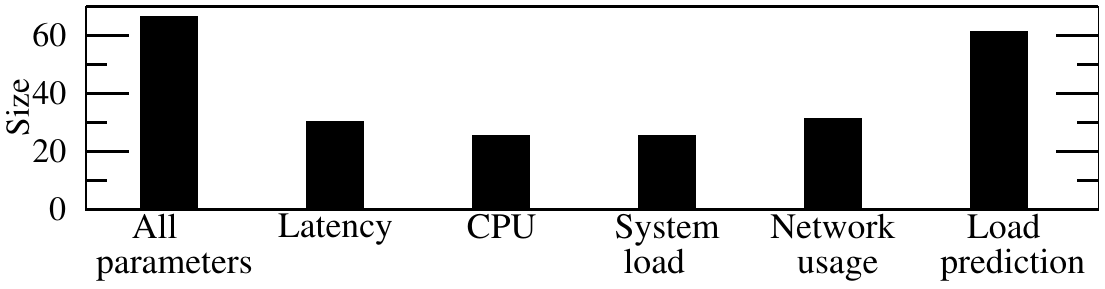}\vspace{-0.1in}
    \caption{Size of the decision tree when allowing splits with only the cluster size
plus one additional parameter}\vspace{-0.1in}
    \label{fig:exp-parameters-splits}
\end{figure}

In order to test the algorithm's ability to partition the state space using different parameters,
as well as to test the reliability of some of the parameters in predicting the incoming load, we
experiment with restricting the parameters with which the algorithm is allowed to partition
the state space. For that purpose, we experiment with training the algorithm from a small
dataset of 1500 experiences, but restricting the parameters with which the algorithm is allowed
to partition the state space to only the size of the cluster plus one additional parameter each
time. The parameters used are the CPU utilization, the one minute
averaged reported system load, the prediction of the incoming load, the network usage
and the average latency.

In Figure \ref{fig:exp-parameters} we present our findings. For all the parameters, the system seems to be able to find a correlation between the given
parameter and the rewards obtained, and starts following the incoming load. Of course,
the performance is significantly worse compared to the default case where all the available
information is provided (Figure \ref{img:allparams}), and thus the training of the model is noticeably slower: In Figures \ref{img:onlylat}, \ref{img:onlycpu}, \ref{img:onlyload}, \ref{img:onlynet} and \ref{img:onlyloadpred} decisions (dotted lines) do not follow workload as smooth as in Figure \ref{img:allparams}. The resulting
size of the decision tree for each individual case reflects this fact, and in most cases
the model ends up having only 20 to 30 states compared to the 66 of the default case (Figure \ref{fig:exp-parameters-splits}).
However, the fact that these correlations exist and can be detected even from a small dataset 
of only 1500 points, reveals the fact that it is possible, using techniques like the ones
described in this work, to exploit these correlations in order to implement policies in
systems with complicated and not very well understood behavior.

\section{Related Work}\label{sec:related}

The proposed system and methodology covers two different areas, namely RL and adaptive resource management. In the former we compare with methods that adaptively partition the state space and in the latter we compare with approaches that manage system resources in a cloud setting.


\textbf{Adaptive State Space Partitioning: } In \cite{chapman}, the authors propose a modification of 
\emph{Q-learning} that uses a decision tree to
generalize over the input. The agent goal is to control a character in a 2D 
video game, where the state is a bit string representing the pixels of the on-screen game representation.
 Since the input consisted of a few hundreds of bits, the state space consisted of more
than $2^{100}$ bits and thus generalization was necessary. 
The proposed algorithm gradually partitioned the state space based on values
of individual bits of the state. A \textit{t} statistic was used to determine if and with 
what bit a state needs to be split.

While this approach succeeded at reducing a very large state space to a manageable number of states,
its applications are limited since it requires that the parameters of the system can only take
two possible values, 0 and 1. At the same time, the information used to perform this partitioning
was thrown as soon as a split was performed, thus wasting valuable information and slowing down the
training.

In \cite{qdt}, a \emph{Q-learning} algorithm that uses a decision tree to dynamically partition
the state space is proposed. The motivation is to build RL agents for
two applications in the field of robotics where the state space is too large for classical
lookup-table approaches. The algorithm builds a decision tree based on values of parameters of
the input, and maintains a \emph{Q-learning} model on the leaves of the tree. Different criteria are
examined for the node splitting, and the algorithm performance is tested
against lookup-table based approaches as well as neural networks.

This algorithm is more widely applicable than the previous approach, since it applies to any
discrete or continuous state space. However, it still throws away training information each
time a state is split. Additionally, it is not based on a full model, which results in it
requiring a larger set of experiences to train, as well as being unable to accurately evaluate
the value of each experience.

In \cite{mccallum1996reinforcement}, a full model based algorithm using a decision tree to partition
the state space is proposed. The algorithm is called \textit{U Tree}, and is able to work strictly
on discrete state spaces. An extension that works on continuous state
spaces is proposed in \cite{continuousutree}, called \textit{Continuous U Tree} which
is split into two phases. During the
\textit{Data Gathering} phase, the MDP states remain unchanged, but experience
tuples are stored for future use. During the \textit{Processing Phase}, the stored experiences
are used to determine the states that need to be split.
Once the new states have been decided, the stored experiences are used to
calculate the transition and reward function for the new set of states, and the values
of the states and Q-states are calculated. The algorithm continuously alternates between the
two phases, periodically extending the decision tree and globally recalculating the current
status of the MDP.

This splitting strategy however proved to not perform as well as splitting the initial state of each
experience, both in terms of training and computational efficiency. Also, the splitting criterion
used was a combination of the Kolmogorov-Smirnov and the \emph{Q-value} tests on multiple points,
which was outperformed by other splitting criteria such as the Mann Whitney U test in combination
with the Parameter test.


\textbf{Adaptive Resource Management}: Elastic resource scaling is typically employed in a cloud setting to regulate resource size and type according to observed workload. Amazon's autoscaling service \cite{aws_autoscale} was one of the first auto scaling services offered through its AWS cloud. It employs simple threshold based rules or scheduled actions based on a timetable to regulate infrastructural resources (e.g., if mean CPU usage is above 40\% then add a new VM). Similar simple rule-based elasticity offerings have been implemented by other cloud providers: Google's cloud platform autoscaler \cite{google_autoscaling}, Rackspace's Auto Scale \cite{rackspace_autoscale},  Microsoft Azure's Autoscale \cite{azure_autoscale} and IBM's Softlayer auto scale \cite{softlayer_autoscale} are a representative small subset of the autoscaling services offered by major cloud vendors. Rule based techniques have been proposed in other systems as well. CloudScale \cite{shen2011cloudscale} employs thresholds to meet user defined SLAs, while it focuses on accurate predictions. Lim et al \cite{lim2010automated} set thresholds to aggregated CPU usage and response time in order to regulate the size of an HDFS cluster. Both ElasTraS \cite{das2013elastras} and SCADS \cite{trushkowsky2011scads}  also employ a rule based approach to decide the former a DB migration command and the latter a distributed storage system scaling action. The same holds for the AGILE system \cite{nguyen2013agile}. Although rule-based approaches are easy to implement and model, they require specific lower level knowledge of both the correct parameters and the respective threshold values, limiting their broader applicability.


On the contrary, systems that employ RL or similar approaches to dynamically manage application resources allow the user to set higher level policies, like, for instance, ``minimize cost and maximize query throughput \cite{tiramola1,rao2009vconf}''.

In \cite{rao2009vconf} a model-based RL algorithm to 
automatically select optimal configuration settings for clusters of VMs is proposed. To tackle
high dimensionality, an environmental model is used to generate experiences
for previously unseen state-action pairs, thus reducing the amount of required training steps. 
The model is represented by a Neural Network that predicts
the expected rewards for the execution of unseen actions based on the real recorded experiences.
This approach allows the algorithm to handle a three dimensional state space, improving the
performance compared to traditional RL solutions. However, since the partitioning
of the state space is still performed following a uniform manner with equally sized intervals, it is impossible for this solution to scale up to 10 or more dimensions. 

Under a similar setting, the authors in \cite{bu2009reinforcement} 
utilize RL techniques to design a system that 
could select optimal configuration settings for online web systems.
In order to handle the large state space resulting from the different
combinations of configuration options, the parameters are grouped based
on their similarity to form a smaller state space. The performance
of the system is sampled for those few representative states and
the performance for the rest of the states was estimated using polynomial
regression. Consequently, the complete model is trained off-line using
this performance estimation. A number of different training sets
are used for the offline training, and the model is retrained with
a different dataset whenever the performance of the system diverged
from the model predictions. Once more, even though these
techniques manage to accelerate the training process, the partitioning
of the state space in the final model is still done in a uniform 
manner with equally sized intervals, limiting the application to low dimensional state spaces.

A more drastic RL approach that involves
splitting the system parameters into two groups to reduce the effect of dimensionality 
is adopted by the authors of \cite{bu2013coordinated}. The
first group includes the configuration parameters of the applications, 
while the second
includes those of the cluster. Each group forms a separate state space, and it is optimized by a different
RL model. Since the number of states grows
exponentially with the number of parameters, this results in a big 
reduction in the total size of the model. Despite this,
the model is still impractically large, and a technique called
\textit{simplex-based space reduction} is implemented to further 
narrow down the state space. Instead of training the complete RL
model, only a small number of points are sampled. Those points are 
gradually replaced through a series transformation rules attempting
to find the regions of the state space that provides the best 
performance for the system. Even though such techniques are 
effective at converting the problem at hand into a more manageable one, however, they
can easily be trapped in local minima and offer no guarantees of
convergence to an optimal configuration.


In \cite{tiramola1} the authors present TIRAMOLA, a cloud-enabled open-source framework to perform
automatic resizing of NoSQL clusters according to user-defined policies. The cluster is modeled as
an MDP, in which the states represent different cluster sizes and the actions
resizing decisions that modify that size. To isolate the most relevant experiences, K-means clustering is used and 
the expected reward is calculated using the cluster centroid. In \cite{tiramola2} the authors extend TIRAMOLA 
to identify different workload types. Nevertheless, in both 
works the authors apply an MDP model using a fixed state model both in terms of parameter size and 
grain. This approach imposes a very detailed description of the ``world'' for the algorithm to behave 
optimally, limiting the system's applicability in cases where this is not feasible.

In \cite{barrett2013applying} the authors employ Q-learning, parallelize its execution and apply it on a use case in which an agent regulates the size of a VM cluster according to observed workload. Nevertheless, since the decision making period is one minute and the benefits of parallelization are in the orders of msec, the use case does not significantly benefit from their optimization. PerfEnforce \cite{ortiz2016perfenforce} offers three scaling strategies to guarantee SLAs for analytical workloads over a cloud setup and one of them is based on RL \cite{tiramola1}. This work only employs RL and does not investigate any algorithmic improvements.

\section{Discussion}\label{sec:discussion}

The performance of the decision tree based models is surprisingly good compared to their traditional RL counterparts. This fact is not only due to the decision tree's ability to create an efficient partitioning of the state space, but also because of the fact that these models adaptively increase their model size as more data become available. This allows them to train quickly at the start of the process, but still gradually increase their size to keep up with larger models as more data were gathered. Additionally, since they do not require      a predefined state space configuration, they can be used in different types of scenarios with the same settings. 

    The splitting criteria that are based on statistical tests are very efficient in distinguishing real correlations from random noise.
    However, in order to achieve this, the error margin needs to be set much lower than the
    typical value of 0.05, depending on the statistical test and splitting criterion used. 
    Of course, since all the criteria detect correlation between the efficiency
    of actions and the values of certain parameters, one still needs to be aware of
    situations where there are temporary correlations between certain parameters
    of the system and its performance. As long as these correlations hold, partitioning
    the state space based on those parameters may not cause a problem, but if these
    correlations suddenly break, the model may stop behaving optimally. For this
    reason, the parameters with which the model is allowed to partition the
    state space is an important decision that needs to be made very carefully.

    The combination of the Mann Whitney U test with the Parameter test splitting criterion
    achieves the best performance among all the tested statistical criteria 
     in both the simulations and the real experiments.
    For the Q-value test in particular, where it is possible to consider 
    multiple splitting points per parameter, attempting to split on only the median achieves
    better results than allowing multiple options, and at the same time produces smaller
    decision trees.

    The fine grained splitting and retraining implemented mechanism 
    performs better than the one used in \cite{continuousutree}, while at
    the same time being more computationally
    efficient. More complicated splitting strategies, like delaying the beginning of the
    splits or periodically resetting the decision tree do not manage to improve performance.
    However, when allowing multiple splitting points per parameter, the latter does not fall
    too far behind the default strategy, and thus could potentially be used to correct
    mistakes caused by misleading data at the start of the training.

    In terms of computational efficiency, maintaining the training data to retrain the new states
    and performing tests on them to decide splits
    does have a considerable effect on the running time, when comparing models with
    approximately equal numbers of states. However, often the decision tree based models
    manage to achieve better performance using a significantly smaller number of states.
    Moreover, in the case of the full-model based approaches, if an update algorithm
    such as prioritized sweeping or value iteration is used to update the values of the
    states and Q-states, the running time is dominated by the performance of the update
    algorithm. As a result, since the running time of these algorithms depends on the
    number of states of the model, decision tree models ended up running faster.

    In the context of cloud computing, where there is generally a lot of computational
    power available and a lot of time between decisions to perform calculations, the
    running time of the algorithms was completely trivial, and the only concern
    was the time needed to perform the initial training from a very large dataset. 
    However, if needed, implementing the algorithms in a statically typed, compiled
    programming language (like C) and using prioritized sweeping as the update algorithm
    makes the training time trivial even in those cases. Of course, in scenarios
    where the computational and energy efficiency is critical (for example when
    controlling mobile robots), Q-Learning provides by far the fastest running time
    and lowest memory requirement,
    at the cost of being the least accurate of the approaches.

    Even though decision tree based algorithms have the potential to model the state
    space with zero knowledge of its topology by starting from a single state, providing
    a little information in the form of a small number of initial states significantly improve
    performance. This is not surprising, since mistakes in the structure of the
    decision tree are much more expensive the closer they are to the root.
    Despite the fact that in most experiments the decision trees used begin as a single
    state, we believe that if applied in practical problems a small configuration
    of starting states should always be used, whenever that kind of information
    is available. In the case of performing elasticity decisions for distributed 
    databases, this kind of information is almost always available, since the current size
    of the cluster and the incoming load are always expected to be a deciding factor in
    the decisions. Therefore, these two parameters should be used
    to create a small initial partitioning of the state space. From that point on, 
    the decision tree algorithm can be used to further partition the state space and
    capture more complicated behaviors of the system that are not obvious beforehand.

\section{Conclusions} \label{sec:conclusions}


In this paper we presented \emph{MDP\_DT}, a reinforcement learning algorithm that adaptively partitions the state space utilizing novel statistical criteria and strategies to perform accurate splits without losing already collected experiences. We calibrated the algorithm's parameters utilizing a simulation environment and we experimentally evaluated \emph{MDP\_DT}'s performance in a real cluster deployment where we elastically scaled a shared-nothing NoSQL database cluster. \emph{MDP\_DT} was able to identify and create only the relevant partitions among tens of parameters, enabling it to take accurate and fast decisions during the NoSQL scaling process over complex not-encountered workloads with minimal initial knowledge, compared to model-free and static algorithms.


\bibliographystyle{IEEEtran}
\bibliography{bibliografia}

\end{document}